\documentclass[aps,preprint,nofootinbib,preprintnumbers,eqsecnum]{revtex4-1}

\usepackage[usenames, dvipsnames]{color}

\newcommand{\nb}[1]{\color{blue}}

\newcommand{\hl}[1]{\color{magenta}}

\usepackage[
	  pagebackref=false,
	  colorlinks=true,
      linkcolor=blue,
      urlcolor=blue,
      filecolor=black,
      citecolor=red,
      pdfstartview=FitV,
      pdftitle={},
        pdfauthor={},
        pdfsubject={},
        pdfkeywords={},
        pdfpagemode=None,
        bookmarksopen=true
      ]{hyperref}

\usepackage[normalem]{ulem}
\usepackage{amsmath}
\usepackage{enumerate}
\usepackage{amsfonts}
\usepackage{epsfig}
\usepackage{mathbbol}

\def\Tr{\mathop{\rm Tr}}

\newcommand\half{{\ensuremath{\frac{1}{2}}}}

\newcommand\field[1]{{\ensuremath{\mathbb{{#1}}}}}

\newcommand\vev[1]{{\ensuremath{\left\langle{#1}\right\rangle}}}

\newcommand\ket[1]{\ensuremath{\lvert{#1}\rangle}}

\newcommand{\HH}{\field{H}}

\newcommand{\OO}{\field{O}}

\newcommand{\RR}{\field{R}}

\newcommand{\be}{\begin{equation}}
\newcommand{\ee}{\end{equation}}
\newcommand{\bea}{\begin{eqnarray}}
\newcommand{\eea}{\end{eqnarray}}
\newcommand{\bega}{\begin{gather}}
\newcommand{\eega}{\end{gather}}

\newcommand{\bi}{\begin{itemize}}
\newcommand{\ei}{\end{itemize}}
\newcommand{\ben}{\begin{enumerate}}
\newcommand{\een}{\end{enumerate}}
\newcommand{\bca}{\begin{cases}}
\newcommand{\eca}{\end{cases}}
\newcommand{\bln}{\begin{align}}
\newcommand{\eln}{\end{align}}
\newcommand{\bst}{\begin{split}}
\newcommand{\est}{\end{split}}
\def\ie{\begin{equation}\begin{aligned}}
\def\fe{\end{aligned}\end{equation}}
\newcommand{\bma}{\le(\begin{matrix}}
\newcommand{\ema}{\end{matrix}\ri)}

\def\b{{\beta}}
\newcommand\ep{\epsilon}

\newcommand\lam{\lambda}

\newcommand\om{\omega}
\newcommand\Om{\Omega}

\newcommand\ga{{\ensuremath{{\gamma}}}}
\newcommand\Ga{{\ensuremath{{\Gamma}}}}
\newcommand\de{{\ensuremath{{\delta}}}}
\newcommand\De{{\ensuremath{{\Delta}}}}

\newcommand\da{{\dagger}}

\newcommand\Th{{\Theta}}

\newcommand\ov{\over}
\newcommand\ha{{\half}}

\def\le{\left}
\def\ri{\right}

\newcommand\sA{{\ensuremath{{\mathcal A}}}}
\newcommand\sB{{\ensuremath{{\mathcal B}}}}
\newcommand\sC{{\ensuremath{{\mathcal C}}}}

\newcommand\sH{{\ensuremath{{\mathcal H}}}}
\newcommand\sK{{\ensuremath{{\mathcal K}}}}

\newcommand\sM{{\ensuremath{{\mathcal M}}}}
\newcommand\sN{{\ensuremath{{\mathcal N}}}}
\newcommand\sO{{\ensuremath{{\mathcal O}}}}

\newcommand\sR{{\mathcal R}}
\newcommand\sS{{\mathcal S}}
\newcommand\sT{{\mathcal T}}
\newcommand\sW{{\mathcal W}}
\newcommand\sX{{\mathcal X}}
\newcommand\sY{{\mathcal Y}}

\newcommand{\bid}{\mathbf{1}}

\newcommand{\wt}{\widetilde}

\begin{document}

\title{Towards a holographic description of closed universes}

\preprint{MIT-CTP/5923}

\author{Hong Liu}
\affiliation{MIT Center for Theoretical Physics---a Leinweber Institute,\\ 
Massachusetts
Institute of Technology, \\
77 Massachusetts Ave.,  Cambridge, MA 02139 }

\begin{abstract}

 \noindent

We study baby closed universes in AdS, focusing on the Antonini-Sasieta-Swingle (AS$^2$) cosmology, which arises in the gravitational description of partially entangled thermal states (PETS), as well as the classical example of Maldacena-Maoz (MM). We show that the algebraic formulation of AdS/CFT---and in particular the subregion-subalgebra duality---provides a natural framework for describing such universes within the standard AdS/CFT setting, phrased in terms of an operator algebra $\sM$ and a state $\omega$ on that algebra, with no need to introduce observers. The algebra encodes all physical operations in the closed universe, and, in principle, all physical observables are obtainable from the pair $(\sM,\omega)$.  Along the way, we propose a resolution to a puzzle raised by Antonini and Rath (AR) and reinforced by recent arguments of Engelhardt and Gesteau and a no-go theorem of Gesteau: that a semiclassical baby universe in the AS$^2$ cosmology cannot be understood from the boundary in the usual manner. Our analysis motivates an averaged large-$N$ limit as part of the AdS/CFT dictionary and points toward a unified treatment of spacetimes of all asymptotics in terms of operator algebras. Finally, our boundary descriptions of closed universes indicate that for small but finite $G_N$ there should exist a semiclassical description of a macroscopic closed universe, rather than a one-dimensional Hilbert space, and we discuss weaknesses in arguments favoring the latter.

\end{abstract}

\today

\maketitle

\tableofcontents

\section{Introduction}

Our current understanding of quantum gravity is tied to the asymptotic structure of spacetime. For asymptotically anti-de Sitter (AdS) spacetimes with a time-like boundary, there exists a precise holographic description in terms of a conformal field theory (CFT) on the boundary. By contrast, for a closed universe or an asymptotically flat spacetime, no equally sharp formulation is known, due to the absence of a suitable boundary on which to define a quantum system. This reliance on asymptotic structure presents a fundamental barrier to extending our understanding of quantum gravity beyond the AdS setting.

Here we advocate an approach that treats all asymptotic structures in a unified manner, extending the algebraic formulation of quantum field theory in curved spacetime (see, e.g.,~\cite{HolWal14} for a review). In this formulation, a quantum gravity system (with a finite $G_N$) is specified by an abstract $*$-algebra $\mathcal{A}$. States are linear maps $\omega:\mathcal{A}\to\mathbb{C}$ satisfying positivity conditions. Spacetimes of differing asymptotic structure correspond to distinct states on the same algebra. Given a state $\omega$ that specifies an asymptotic structure, the Gelfand-Naimark-Segal~(GNS) construction furnishes a Hilbert space $\sH_\omega$, within which the full physics of spacetimes sharing that asymptotic structure is realized. The Hilbert space $\sH_\omega$ carries a representation of the algebra, denoted by $\pi_\omega(\sA)$.

For example, consider type IIB superstring theory, and suppose that its non-perturbative completion could be described by an abstract algebra $\sA_{\rm IIB}$. A background such as $\mathrm{AdS}_5 \times S_5$ could then be interpreted as corresponding to a state $\omega_{\mathrm{AdS}_5 \times S_5}$ on $\sA_{\rm IIB}$, with the associated GNS Hilbert space identified with the Hilbert space of $\mathcal{N}=4$ Super-Yang-Mills (SYM) theory. Similarly, $\mathrm{AdS}_3 \times S_3 \times K3$ would correspond to another state, with the resulting GNS Hilbert space identified with that of the (deformed) symmetric orbifold theory. Each realization of the AdS/CFT correspondence involving type IIB superstring theory would thus correspond to a distinct state on the same algebra $\sA_{\rm IIB}$, with different boundary CFTs providing different representations $\pi_\om (\sA_{\rm IIB})$ of this algebra. Ten-dimensional Minkowski spacetime and potential de Sitter vacua may be interpreted in the same way, although we still lack a holographic description for them.
In this framework, all solutions of type IIB string theory could be understood in a unified manner, independent of their asymptotic structure.


At present, we do not have a background-independent definition of $\sA_{\mathrm{IIB}}$, nor a complete characterization of the allowed states. String field theory (see, e.g.,~\cite{Erb21,SenZwi24} for recent reviews) represents an important step in this direction. In the present algebraic language, string field theory yields, for a chosen background state $\omega$, the Hilbert space $\sH_\omega$ together with the background-dependent algebra $\pi_\omega(\sA_{\mathrm{IIB}})$. Furthermore, string field theory can in principle be used to systematically identify further consistent states and the corresponding algebras. 
See also~\cite{Wit23b} for a recent proposal to define a background-independent algebra in quantum gravity using an algebra of operators along an observer's worldline.

Within this framework, Hilbert space is not fundamental but rather a derived construct, introduced only to furnish representations of certain algebras. The familiar semiclassical geometric description---and with it the standard formulation of quantum mechanics---emerges from these algebraic structures in the limit $G_N \to 0$, with the relevant physical scales held fixed.

In this paper, we provide modest support for the general framework outlined above by presenting a holographic description, in algebraic terms, of certain closed universes that arise within AdS/CFT in the $G_N \to 0$ limit. A parallel algebraic description can be developed for asymptotically flat spacetimes, which will be presented elsewhere.\footnote{Work in progress with Daiming Zhang.}



More concretely, we consider the baby closed universe of Antonini-Sasieta-Swingle~(AS$^2$)~\cite{AntSas23}, which arises in the gravitational description of partially entangled thermal 
states~(PETS)~\cite{GoeLam18}, as well as the classical example of Maldacena-Maoz (MM)~\cite{MalMao04}. 
 We show that the algebraic formulation of AdS/CFT, and in particular the subregion-subalgebra duality~\cite{LeuLiu21b,LeuLiu22}, provides a natural conceptual framework for describing these universes within the standard AdS/CFT setting, phrased in terms of an operator algebra $\sM$ and a state $\omega$ on that algebra. 
The algebra encodes all physical operations in the closed universe, and, in principle, all bulk observables can be extracted from the data of the state together with the algebra. There is no need to introduce any observers by hand,\footnote{See e.g.,~\cite{ChaLon22,Gom23c,Wit23b,KudLeu24,ChePen24,KolLiu24} for recent discussions of operator algebras in a closed universe by introducing observers.} although we do expect physics should be relational. 

Along the way, we propose a resolution to a puzzle raised by Antonini and Rath (AR)~\cite{AntRat24}, who argued that 
there seems to be two possible bulk descriptions for the same state, one with the baby universe and one without.\footnote{AR considered three alternative possibilities: (i) AdS/CFT cannot provide a complete description of the bulk, (ii) the baby universe fails to admit a semiclassical description, or (iii) the boundary description of the baby universe  requires ensemble averages over theories.}  See also~\cite{EngGes25,EngGes25b,Hig25,AntRat25} for related discussions. This puzzle was further sharpened by Engelhardt and Gesteau~\cite{EngGes25}, and in particular Gesteau~\cite{Ges25}  established a no-go theorem ruling out a semi-classical description of the baby universe from the boundary under reasonable assumptions.\footnote{The main assumption is that correlation functions of single-trace operators have a well-defined $N \to \infty$ in the PETS.}  
We show that the baby universe can, in fact, be described within the standard AdS framework using boundary algebras---provided:  
\ben  
\item one postulates, in the boundary CFT, $N$-dependent oscillatory behavior in the matrix elements of the heavy operator defining PETS between light states,  
\item and one defines the large-$N$ limit appropriately by averaging over $N$.  
\een

That an averaged large-$N$ limit may be needed when comparing the boundary CFT with semiclassical bulk gravity was discussed in~\cite{SchWit22} in the context of black hole states, as an alternative to ensemble averages over different theories. Our discussion, which is about vacuum-sector states, reinforces this proposal and suggests that it should be regarded as an essential component of the AdS/CFT dictionary.

Another simple but important observation from our discussion is that, in the case of AS$^2$ cosmology, each of the right and left boundaries possesses a causal wedge but lacks a well-defined entanglement wedge. This feature should be generic for entanglement of order $O(G_N^0)$, in contrast to entanglement of order $O(1/G_N)$.

Our boundary descriptions for the closed universes in~\cite{AntSas23,MalMao04}  suggest that for small but finite $G_N$, there exists a semiclassical description of a macroscopic closed universe, rather than a trivial one-dimensional Hilbert space (see e.g.,~\cite{McNVaf20,MarMax20,UsaWang24,UsaZha24,HarUsa25,AbdAnt25} for recent discussions). We also discuss weaknesses in the arguments that favor a one-dimensional Hilbert space.

\begin{figure}[h]
        \centering
			\includegraphics[width=12cm]{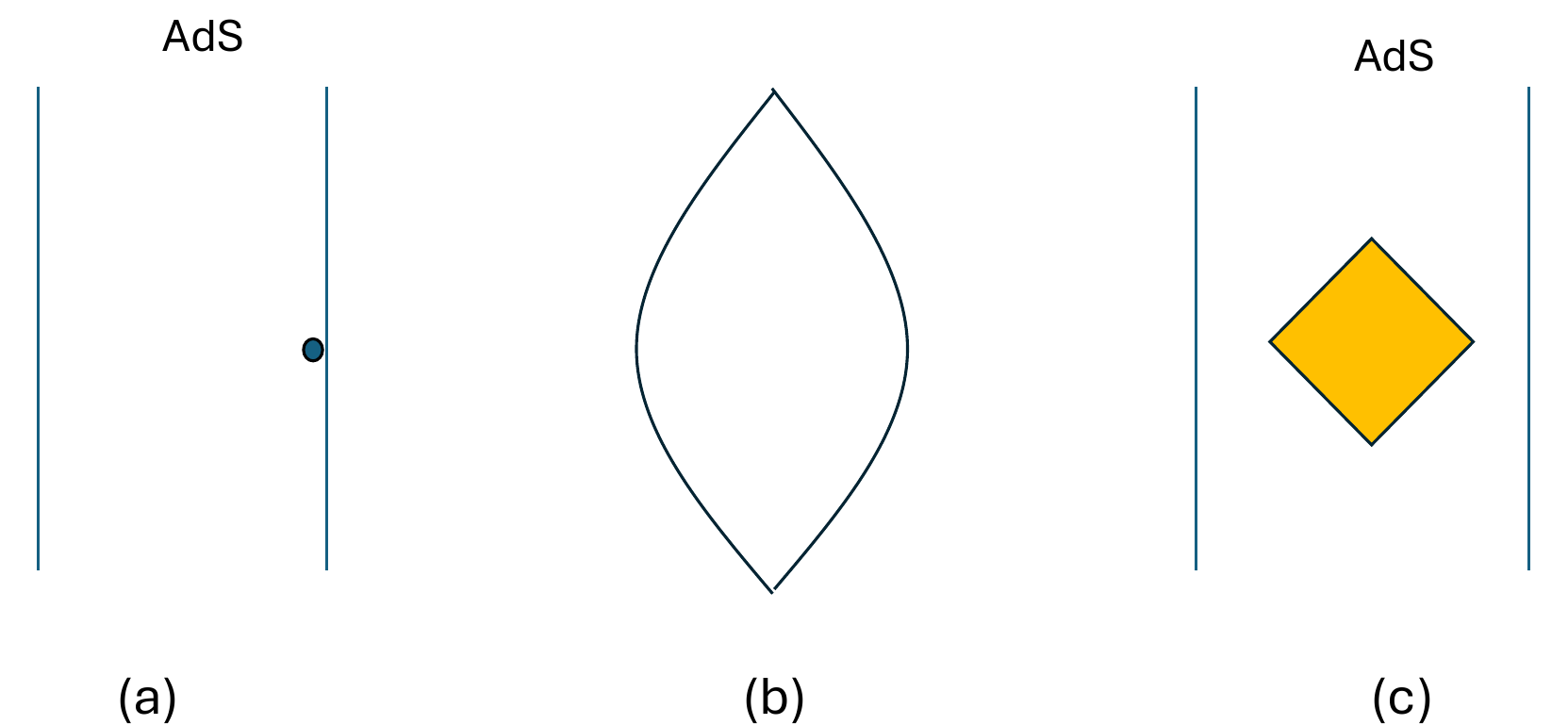}
                \caption[  ]
        {\small (a) Global AdS.  (b) A closed universe. (c) A causal diamond region (shaded region) in global AdS that does not touch the boundary.
        The algebra for (a) contains ``local'' operators that arise from the boundary limit of bulk operators (represented by a dot in the plot), which are absent in those for (b) and (c).  The algebras for (a) and (b) are type I, while that for (c) is type III$_1$. 
        } 
\label{fig:com}
\end{figure}

 The algebras obtained here for closed universes are not the background-independent algebras mentioned earlier, but rather ones constructed from the background-dependent boundary CFT. Nevertheless, this discussion shows that closed universes can be given a holographic description, even in the absence of a boundary, providing a unified treatment of AdS and closed universes within a common algebraic framework. The absence of a boundary in a closed universe is reflected in the structure of its operator algebra, through its ``good ultraviolet'' behavior, in contrast to the algebras associated with asymptotically AdS spacetimes~(see Fig.~\ref{fig:com} for a comparison of algebras describe different bulk systems). In other words, the defining feature of bulk systems with boundaries is the presence of a special subset of ``local'' operators, obtained from bulk fields through the boundary limit via the extrapolate dictionary. These descriptions of closed universes point toward a form of holography in which operator algebras, rather than boundaries, provide the essential structure.

%



The plan of the paper is as follows. In Sec.~\ref{sec:duality}, we present a brief review of the algebraic formulation of the AdS/CFT duality in the large-$N$ limit.  In Sec.~\ref{sec:ASS}, we discuss the boundary description of the baby universe of AS$^2$~\cite{AntSas23}. In Sec.~\ref{sec:MM}, we outline a holographic description of the closed universe of MM~\cite{MalMao04}.
Sec.~\ref{sec:Conc} begins with a summary of the paper and an outlook on future directions. We then revisit recent arguments suggesting that a closed universe admits only a one-dimensional Hilbert space.

\smallskip

\noindent {\bf Notations and conventions} 

$\bullet\,$ $\sB (\sH)$ denotes the set of bounded operators on a Hilbert space $\sH$.

$\bullet\,$ Throughout the paper we consider a gravitational system in AdS$_{d+1}$ described by a $d$-dimensional boundary CFT$_d$ on $\RR \times S_{d-1}$. We assume that the CFT has a parameter $N$, related to the bulk Newton constant by $G_N \propto 1/N^2$.
The AdS radius is set to unity. 

\medskip 

\noindent {\bf Note added:} After completing this work, we learned that Jonah Kudler-Flam and Edward Witten are working in a somewhat similar direction on AS$^2$ cosmology.

\section{Algebraic formulation of the AdS/CFT duality in the large-$N$ limit} 
\label{sec:duality}

Here we briefly review the essential elements of the algebraic formulation of the AdS/CFT duality in the large-$N$ limit~\cite{LeuLiu21b,LeuLiu22} that will be used to describe closed universes in subsequent sections.

Consider a pure state $\ket{\Psi}$ in the boundary system that admits a bulk description in terms of a classical geometry in the large-$N$ limit---we will refer to such a state as a semi-classical state.
By definition, a semi-classical state has a  large-$N$ limit; in particular, $\ket{\Psi}$ can be regarded as the $N \to \infty$ limit of a sequence of states ${\ket{\Psi_N}}$, one for each finite $N$.

We say that an operator $A$ has a well-defined large $N$ limit in a semi-classical state $\ket{\Psi}$ if $A$ can be regarded as the $N \to \infty$ limit of a sequence of operators $\{\ket{A_N}\}$ with\footnote{As will be explained in more detail in Sec.~\ref{sec:largeN}, the definition of the large $N$ limit of~\eqref{opD} may require particular care in certain situations.}
\be \label{opD}
\lim_{N \to \infty} \vev{\Psi_N|A_N| \Psi_N} < \infty \ .
\ee
We denote by $\sA_{\Psi}$ the vector space of operators that admit a well-defined large-$N$ limit in the state $\ket{\Psi}$.
 We further assume that $\sA_{\Psi}$ forms a $*$-algebra; that is, if $A \in \sA_\Psi$ then $A^\da \in \sA_{\Psi}$, and if $A, B \in \sA_\Psi$ then $AB \in \sA_{\Psi}$.
 

We define single-trace operators in the boundary CFT as the boundary limits of bulk elementary field operators via the extrapolation dictionary. By construction, the algebra generated by these single-trace operators survives the large-$N$ limit for all semiclassical states, and therefore constitutes a universal subalgebra of $\sA_\Psi$. We will denote it as 
\be \label{SingOp}
\sS \equiv 
\text{$*$-algebra generated by single-trace operators} \subseteq \sA_\Psi \ .
\ee
$\sA_\Psi$ may coincide with $\sS$, but can also contain additional operators whose existence depends on the specific semi-classical state $\ket{\Psi}$. 

Expectation values in $\ket{\Psi}$ in the large $N$ limit define a state $\om_\Psi$ on the algebras $\sA_\Psi$ and $\sS$, 
\be\label{Dsta}
\om_{\Psi} (A) = \vev{\Psi|A|\Psi} , \quad A \in \sA_\Psi \ .
\ee
The action of $\om_\Psi$ on $\sA_\Psi$ can be used to construct a GNS Hilbert space $\sH_\Psi^{(\rm GNS)}$, which can be heuristically thought of as the Hilbert space of ``low-energy'' excitations around $\ket{\Psi}$ from acting elements of $\sA_\Psi$ on $\ket{\Psi}$. 
We will denote the representation of $\sA_\Psi$ on $\sH_\Psi^{(\rm GNS)}$ as $\pi_\Psi (\sA_\Psi)$. 
At finite $N$, $\ket{\Psi}$ is a pure state and gives a complete description of the system. We thus expect $\om_\Psi$ to be a pure state on  $\sA_\Psi$, which implies
\be \label{xrel}
\pi_\Psi (\sA_\Psi) = \sB (\sH_\Psi^{(\rm GNS)}) \equiv \sX\ .
\ee

We denote the bulk gravity solution (which includes the spacetime geometry as well as possible matter configurations) dual to a semi-classical state $\ket{\Psi}$ as $g_\Psi$. Quantizing small bulk excitations around $g_\Psi$ leads to a Fock space which we will denote as $\sH^{\rm (Fock)}_\Psi $. 
The AdS/CFT duality implies that we should have the identification 
\be \label{imid}
\sH^{\rm (Fock)}_\Psi = \sH_\Psi^{(\rm GNS)} , \quad \sB (\sH^{\rm (Fock)}_\Psi ) = \sX  \ .
\ee
The second equation of~\eqref{imid} in particular implies that there should be a one-to-one correspondence between bulk subalgebras of $\sB (\sH^{\rm (Fock)}_\Psi )$ and boundary subalgebras of $\sX$, which leads to the subregion-subalgebra duality~\cite{LeuLiu22}.

Denote by $\sY \equiv \pi_\Psi(\sS)$ the representation of the single-trace algebra $\sS$ on $\sH_\Psi^{(\rm GNS)}$.
According to subregion-subalgebra duality, $\sY$ is dual to the causal wedge of the full boundary.
In the case where $\sS$ is a proper subset of $\sA_\Psi$, i.e., $\sY$ is proper set of $\sX$, the duality implies that the full bulk geometry extends beyond the causal wedge of the full boundary. In particular, there should exist regions that cannot be reached from the boundary by light rays---for example, the interior of a black hole or a baby universe. Such regions can be probed using the commutant $\sY'$.

\section{Holographic Description of Baby Universes in AS$^2$ Cosmology} \label{sec:ASS}

In this section, we show that the algebraic formulation of the AdS/CFT duality, reviewed above, leads to a natural and powerful way to understand how the baby universe is encoded in the boundary system. More specifically, by the subregion-subalgebra duality, the baby universe should be described by a certain subalgebra of the boundary theory in the large-$N$ limit.

\subsection{Review of the setup} \label{sec:rev}

Take two copies of the boundary CFT, denoted CFT$_R$ and CFT$_L$, with (finite-$N$) Hilbert space $\sH_R \otimes \sH_L$. A partially entangled thermal state (PETS) is defined as~\cite{GoeLam18}:
\bea\label{PETS} 
\ket{\Psi_\OO^{(\b_R, \b_L)}} 
& =& {1 \ov \sqrt{Z}}  \sum_{m,n} e^{- \ha \b_L  E_m} e^{- \ha \b_R E_n} \OO_{nm} \ket{n}_R  \ket{\tilde m}_L  \\
& =&  {1 \ov \sqrt{Z}} \OO_R \le({i \ov 2} \b_R \ri) \ket{\Psi_\b} , \quad \b = \b_R + \b_L ,
\label{PETS1} 
\eea
where $\OO$ is a heavy operator of dimension $\De_\OO \sim O(N^2)$, uniformly smeared over the spatial directions, and $\OO_{nm} = \langle n|\OO|m\rangle$ are its matrix elements in the energy eigenbasis ${\ket{m}}$ with corresponding eigenvalues $E_m$. The states $\ket{\tilde m} = \Th \ket{m}$ are defined using a time-reversal operator $\Th$ (such as $\sC\sR\sT$).  $\ket{\Psi_\b}$ denotes the thermofield double (TFD) state with inverse temperature $\b$
\be\label{tfd}
\ket{\Psi_\b} =  {1 \ov \sqrt{Z_\b}} \sum_m e^{-\ha \b E_m} \ket{m}_R \ket{\tilde m}_L ,
\ee
where $Z_\b$ denotes the thermal partition function of inverse temperature $\b$. 
The normalization factor $Z$ in~\eqref{PETS}--\eqref{PETS1} is given by
\be \label{parti}
Z = \vev{\Psi_\b|\OO^\da (- i \b_R/2) \OO (i \b_R/2)|\Psi_\b} \equiv \vev{\OO^\da (- i \b_R/2) \OO (i \b_R/2)}_\b
\ .
\ee
We also introduce 
\be 
\mu_\OO \equiv {\De_\OO \ov N^2} 
\ee
which measures backreaction of the mass shell resulting from $\OO$ insertion on the gravity side.\footnote{It is related by an $O(1)$ constant to $G_N m$, with $m$ the mass of the matter shell representing the operator in the gravity description.} 

\begin{figure}[h]
        \centering
			\includegraphics[width=14cm]{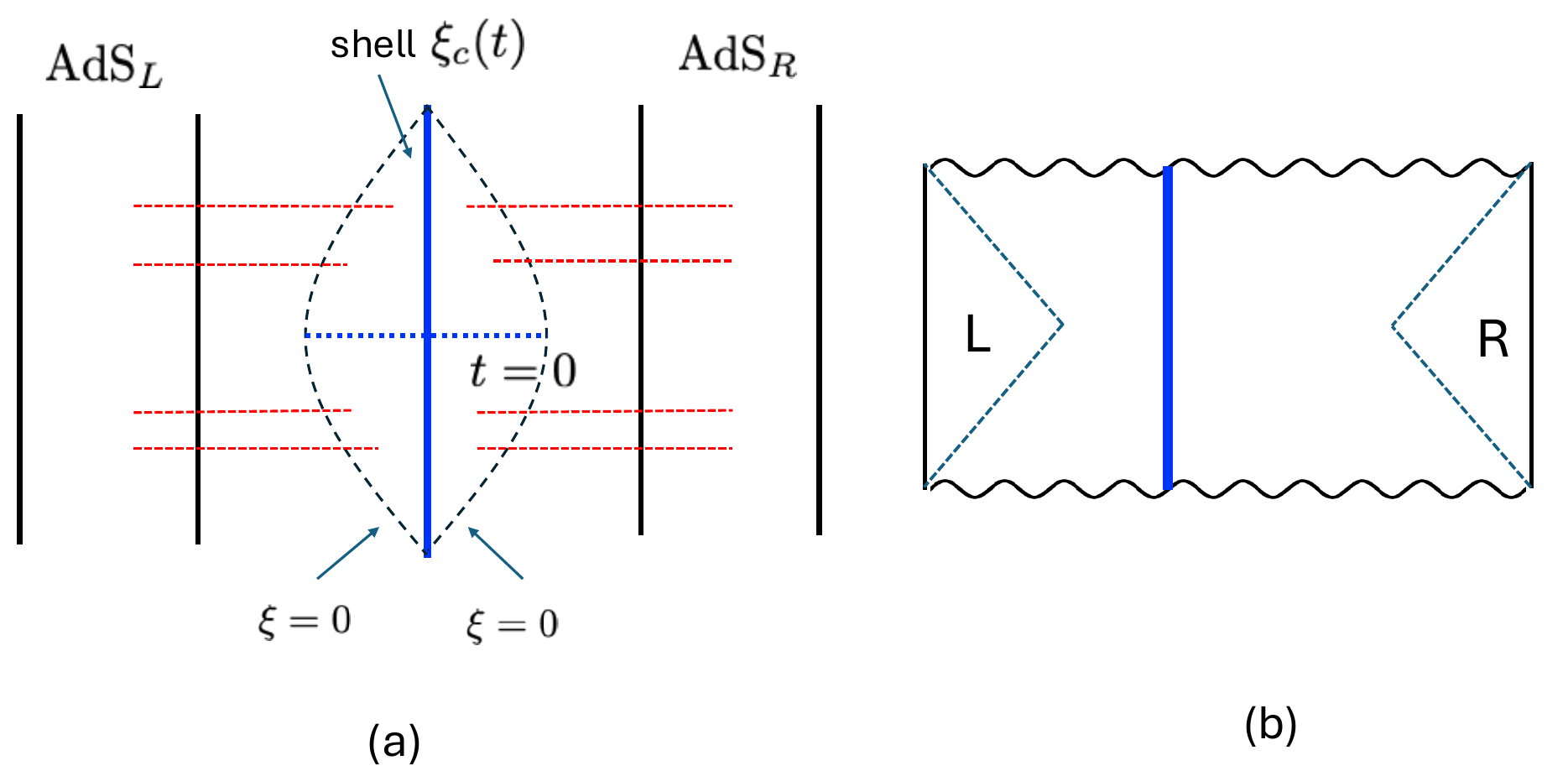}
                \caption[  ]
        {\small Gravity description of PETS~\eqref{PETS}. The insertion of the heavy operator $\OO$ generates a matter shell in the bulk geometry, represented in the plots by thick vertical lines.
(a) In the TAdS-BU phase (for sufficiently low temperature), the geometry consists of two copies of global AdS entangled with a baby universe. The baby universe is constructed by gluing together two copies of~\eqref{einn0} along $\xi_c(t)$, the location of the shell. The center $\xi = 0$ of each patch is shown as a dashed line. The shell position $\xi_0$ on the $t=0$ slice (shown as a dotted blue line) diverges as $\mu_\OO \to \infty$, effectively covering an entire slice of global AdS. Dotted red lines are schematic cartoons for entanglement. 
(b) At sufficiently high temperature, the system is described by a long two-sided black hole.
        } 
\label{fig:baby}
\end{figure}

In the $N \to \infty$ limit, as for the thermofield double state $\ket{\Psi_\b}$, the PETS $\ket{\Psi_\OO^{(\b_R, \b_L)}}$ can exhibit different phases depending on the values of $\b_R, \b_L$. In particular, for sufficiently large $\b_L, \b_R$, it was shown in~\cite{AntSas23}  that~\eqref{PETS} is described on the gravity side by two copies of global AdS entangled with a baby closed universe. See Fig.~\ref{fig:baby}(a).  We will refer to this phase as the {\it thermal AdS-baby universe} (TAdS-BU) phase, to emphasize both its similarity to the usual thermal AdS phase and its key new feature: presence of a baby closed universe. In contrast, for sufficiently small $\b_L, \b_R$, the gravity description is given by a long black hole~(as illustrated in Fig.~\ref{fig:baby}(b)), to which we will refer as the long-BH phase. 

PETS in the TAdS-BU phase provide an exciting opportunity for studying big bang/big crunch closed universes using the AdS/CFT duality. Here is a summary of some main features~\cite{AntSas23}: 

\ben 

\item The baby universe can be constructed by gluing together, along the matter shell, two copies of global AdS cut off at a radial location determined by the shell. A full time slice is compact and topologically a $d$-dimensional sphere. More explicitly, the metric for one copy can be written as
\begin{equation} \label{einn0}
ds^2 = -\cosh^2 \xi  dt^2 + d \xi^2 + \sinh^2 \xi  d \Omega_{d-1}^2, \quad \xi < \xi_c(t) ,
\end{equation}
where $\xi_c(t)$ denotes the location of the shell at which the spacetime is cut off.
We take $\xi_c(t)$ to be symmetric under time reflection about the $t=0$ slice, with $\xi_c(0) \equiv \xi_0$, and monotonically decreasing to zero at some finite $t_c < \tfrac{\pi}{2}$. 

In the limit $\mu_{\mathcal O} \to \infty$, one has $\xi_0 \to \infty$ and $t_c \to \tfrac{\pi}{2}$. In this regime, the $t=0$ slice of a single patch covers a full spatial slice of global AdS, and~\eqref{einn0} approaches a Wheeler-DeWitt patch of global AdS.

\item The energies of the CFT$_R$ and CFT$_L$ are of order $O(N^0)$. The partition function~\eqref{parti} has the general structure 
\be 
Z = e^{- N^2 f_0} F (\b_R, \b_L)
\ee
where $f_0$ is an $O(N^0)$ constant independent of $\b_R, \b_L$, and $F (\b_R, \b_L)$ is an order $O(N^0)$ function of $\b_R, \b_L$ (both $f_0$ and $F$ can depend on $\mu_\OO$).  The story is particularly simple in the $\mu_\OO \to \infty$ limit, where the dominant Euclidean saddle in the computation of $Z$ is essentially two copies of Euclidean thermal AdS, with inverse temperatures $\beta_R$ and $\beta_L$. That is, AdS$_R$ and AdS$_L$ are respectively in thermal states at temperatures $\beta_R^{-1}$ and $\beta_L^{-1}$, purified by the baby universe. The corresponding $Z$ is given by 
\be \label{gravP}
Z = Z_{\b_L} Z_{\b_R} e^{- N^2 f_0}, \quad \mu_\OO \to \infty, 
\ee
where $Z_\b$ denotes the boundary CFT partition function at inverse temperature $\b$. Note that both $Z_{\b_R}$ and $Z_{\b_L}$ are of order $O(N^0)$ for $\b_R, \b_L$ in the TAdS-BU phase. 

\item
In the $N \to \infty$ limit, the entanglement and R\'enyi entropies between CFT$_R$ and CFT$_L$ can be computed on the gravity side using the replica trick and are of order $O(N^0)$.
Their expressions again simplify greatly in the $\mu_\OO \to \infty$ limit. In this regime, the second R\'enyi entropy $S_2^{(R)}$ and von Neumann entropy $S_R$ for CFT$_R$ at the time-reflection symmetric slice $t=0$ are respectively given by\footnote{The expressions below were obtained without averaging over the normalization factor, and do not hold at extremely low temperatures, when $S_{\b}$ and $S_{2, \b}$ are close to $0$. This form will be convenient for our later purpose.}
\be \label{renyiE}
e^{- S_2^{(R)}}= e^{-S_{2, \b_R}} + e^{-S_{2 ,\b_L}} , \quad
e^{-S_R} = e^{-S_{\b_R}} + e^{-S_{\b_L}} ,
\ee
where $S_{2, \b}$ and $S_\b$ are the second R\'enyi entropy and von Neumann entropy for the boundary CFT in a thermal state with inverse temperature $\b$, respectively. 

It is import to stress that at the temperatures under considerations all the quantities in~\eqref{renyiE} are of order $O(N^0)$. In particular, this implies that we should keep both terms on the right hand sides, even if parametrically one term is much larger than the other. 
In other words, it is {\it not} correct to write $S_R= {\rm min} (S_{\b_L}, S_{\b_R})$ conceptually, even though parametrically this may be a good approximation. As we will see in Sec.~\ref{sec:AEW}, this has important implications.

\een 

From now on, for notational simplicity, we will simply write $\ket{\Psi_\OO}$ for $\ket{\Psi_\OO^{(\b_R, \b_L)}}$.


\subsection{Algebraic description of the baby universe: the basic picture}

In this subsection, we give a boundary description of the baby universe following the algebraic framework reviewed in Sec.~\ref{sec:duality}.  

We denote the algebra of operators surviving the large $N$ limit in $\ket{\Psi_\OO}$ as 
\be 
\sA_{\Psi_\OO} = \sA_R \otimes \sA_L, 
\ee
where $\sA_R$ denotes the algebra of operators from CFT$_R$ surviving the large $N$ limit. The algebra of single-trace operators 
is $\sS = \sS_R \otimes \sS_L$. Expectation values in $\ket{\Psi_\OO}$ define a state $\om_{\Psi_\OO}$ on the algebras $\sA_{\Psi_\OO}$ and $\sS$~(recall~\eqref{Dsta}). The GNS Hilbert space resulting from the action of $\om_{\Psi_\OO}$ on $\sA_{\Psi_\OO} $
is denoted as $\sH_{\Psi_\OO}^{(\rm GNS)}$, which is identified with the Fock space $\sH^{\rm (Fock)}_{\Psi_\OO}$ of the bulk gravity theory. The representations of $\sA_{\Psi_\OO} $ and $\sS$ on $\sH_{\Psi_\OO}^{(\rm GNS)}$ are 
\bega 
\sX \equiv \pi_{\Psi_\OO} (\sA_{\Psi_{\OO}}) =  \sX_R \lor \sX_L , \quad 
\sX_R \equiv \pi_{\Psi_\OO} (\sA_R), 
 \\ 
\sY \equiv \pi_{\Psi_\OO} (\sS) =  \sY_R \lor \sY_L,\quad
\sY_R \equiv   \pi_{\Psi_\OO}  (\sS_R) \ .
\end{gather} 
We assume that $\om_{\Psi_\OO}$ is a pure state for  $\sA_{\Psi_\OO} $, which translates into the statement 
\be
\sX= \sB (\sH_{\Psi_\OO}^{(\rm GNS)}) \ .
\ee
When $\sS$ is a proper subset of $\sA_{\Psi_\OO}$, the state $\om_{\Psi_\OO}$ is mixed with respect to $\sS$, and the corresponding algebra $\sY$ has a nontrivial commutant in $\sX$.


In the TAdS-BU phase of PETS~\eqref{PETS}, the bulk Hilbert space has the structure 
\be \label{heg}
\sH^{\rm (Fock)}_{\Psi_\OO} =  \sH_\Om^{(R)} \otimes \sH_\Om^{(L)} \otimes \sH_{\rm BU}^{(\rm Fock)} \ .
\ee
Here $\sH_\Om^{(\rm R,L)}$ (with $\Om$ denoting the vacuum state) denotes the Fock space of the bulk gravity system in the $R,L$ global AdS, which can in turn be identified as the GNS Hilbert space of the boundary CFTs in the vacuum sector.
 $\sH_{\rm BU}^{(\rm Fock)}$ denotes the bulk Hilbert space of the baby universe.

The causal wedge of the full boundary consists of the two copies of global AdS, which gives the identification 
\be \label{heg1} 
\sY_R = \sB (\sH_\Om^{(R)}) , \quad \sY_L = \sB (\sH_\Om^{(L)}), \quad \sY = \sB (\sH_\Om^{(R)}) \otimes \sB (\sH_\Om^{(L)}) \ .
\ee
It is now conceptually simple to describe the baby universe: given~\eqref{heg}, it should be dual to the commutant $\sY'$, i.e., we
can identify 
\be 
\sY' = \sB (\sH_{\rm BU}^{(\rm Fock)}) \ .
\ee

Therefore, understanding whether the baby universe admits a semi-classical regime---and its corresponding boundary description---reduces to analyzing the structure of and $\sX$ and the commutant $\sY'$ inside $\sX$. In particular, the possible existence of a commutant of $\sY$ can be inferred from whether the state $\omega_{\Psi_\OO}$ is mixed with respect to $\sS$.



\subsection{Existence of commutant of the single-trace algebra  and the AR puzzle}\label{sec:comm}

In this subsection, we give boundary arguments that the commutant $\sY'$ of the single-trace algebra $\sY$ is non-empty.

For this purpose, we begin with a puzzle raised in~\cite{AntRat24}, which can be stated as follows. Since the contribution to the partition function $Z$ (equation~\eqref{parti}) is dominated, in the large-$N$ limit, by states of energies of order $O(N^0)$, one should be able to truncate the sum in~\eqref{PETS} to this sector. Accordingly, in the large-$N$ limit, $\ket{\Psi_\OO}$ can be approximated as a state in $\sH_\Omega^{(R)} \otimes \sH_\Omega^{(L)}$, 
\be \label{ps1}
\ket{\Psi_\OO} = \sum_{a,b} \psi_{ab} \ket{a}_R \ket{\tilde b}_L + \cdots 
\ee
where $a$ and $b$ run over energy eigenstates of order $O(N^0)$, and $\cdots$ denotes contributions from states whose energies diverge as $N \to \infty$, with their contributions to the partition function being exponentially suppressed in $N$.

Equation~\eqref{ps1} is in tension with the gravity description, as it represents an entangled pure state in two copies of global AdS, leaving no room for a baby universe Hilbert space. In the algebraic language introduced above, the state of~\eqref{ps1} is a pure state for the single-trace algebra $Y = \sB(\sH_\Omega^{(R)}) \otimes \sB(\sH_\Omega^{(L)})$, and consequently its commutant $\sY'$ is trivial; there is no baby universe. This observation led~\cite{AntRat24} to question whether the baby universe admits a semiclassical description (see also~\cite{EngGes25} for further elaboration), or whether AdS/CFT provides a complete description of bulk quantum gravity .

There is, in fact, a way out. By identifying a loophole in the preceding argument, we will argue that (1) the baby universe can admit a semiclassical description, and (2) AdS/CFT can provide a complete and consistent framework for its description.

In writing down~\eqref{ps1} as a state in $\sH_\Omega^{(R)} \otimes \sH_\Omega^{(L)}$, 
it was implicitly assumed that the coefficients $\psi_{ab}$, which are related to $\OO_{ab}$ as 
\be \label{psab}
\psi_{ab} = {1 \ov \sqrt{Z}}  e^{- \ha \b_L  E_b} e^{- \ha \b_R E_a} \OO_{ab} 
\ee
have a well-defined $N \to \infty$ limit (i.e., are {\it $N$-independent} in the limit). This is indeed a reasonable assumption for the amplitude of $|\psi_{ab}|$, but matrix elements $\OO_{ab}$ can in principle have $N$-dependent phases, which as we will see bleow can have important implications.  

More explicitly, consider the following ansatz\footnote{A similar ansatz was discussed in~\cite{Sas22,AntSas23,AntRat25}, with an important conceptual difference to be discussed below. See also the second-to-last paragraph of Sec.~\ref{sec:largeN} for further remarks on the ansatz.}
\bega \label{newsan}
\OO_{ab} = e^{-\ha N^2 f_0}  f^\ha (E_a, E_b) e^{i N^2 g_{ab}} , 
\quad g_{ab} = g (E_a, E_b), 
\end{gather} 
where $f_0$ is independent of $a,b$, while $f (E_a, E_b) > 0$ and $g (E_a, E_b)$ are  smooth real functions of $E_a$ and $E_b$. All these quantities are of order $O(N^0)$.
While equation~\eqref{newsan} is reminiscent of the ETH ansatz, it is not related to thermalization and the physics involved is different. It should be viewed as a kind of ``hydrodynamic'' approximation: for a heavy operator $\OO$, the detailed structures of low-energy states $\ket{a}$ and $\ket{b}$ are irrelevant, except for their energies. 

The partition function $Z$ depends only on the magnitude of $\OO_{ab}$ 
\bega
Z  = \vev{\OO^\da (-i \b_R/2) \OO (i \b_R/2) }_\b = \sum_{a,b} e^{-\b_L E_a} |\OO_{ba}|^2   e^{-\b_R E_b} \\
=  e^{- \ha N^2 f_0} \sum_{a,b}  e^{-\b_R E_b}  e^{-\b_L E_a}  f (E_b, E_a)  ,
\end{gather} 
and the gravity result for $Z$ can be qualitatively reproduced with an appropriate choice of $f(E_a, E_b)$~\cite{Sas22,AntSas23}. 
For example, the simplest case 
\be 
f (E_a, E_b) =1 \quad \to \quad Z = Z_{\b_L} Z_{\b_R} e^{- N^2 f_0}, 
\ee
reproduces the gravity result~\eqref{gravP} in the $\mu_\OO \to \infty$ limit. This can be intuitively understood as the statement that, in this limit, even the energies of light states become negligible in the magnitude of $\OO_{ab}$. 

Here it is important that the $N$-dependent factor $e^{- \ha N^2 f_0}$ in $|\OO_{ab}|$ has no dependence on $a, b$. Otherwise, it would be incompatible with the gravity result that the energy expectation values for $R$ and $L$ systems are $O(N^0)$.  Consequently, the factor cancels out in~\eqref{psab}. While the phase in~\eqref{newsan} is not probed by the partition function or energy expectation values, it has important consequences for correlation functions in the state $\ket{\Psi_\OO}$ and entropies. 

Consider the simplest case $f (E_a , E_b) =1$ and assume 
\bega
\quad g_{ab} \neq g_{cd} \;\; \text{except for } a=c, \; b=d  \ .
\label{newsan1}
\end{gather} 
Two-point functions of single-trace operators in CFT$_R$ in $\ket{\Psi_\OO}$ can then be written as
\bea \label{grr}
G_{RR} &=& \vev{\Psi_\OO|\sO_R (t_1) \sO_R( t_2)|\Psi_\OO}
= \vev{\OO^\da(-i \b_R/2) \sO(t_1) \sO(t_2) \OO (i \b_R/2)}_\b \\
&=&  {1 \ov Z} \sum_{a,b,c,d} e^{-\b_L E_a } \OO^*_{ba}  e^{-(\b_R /2 - i t_1) E_b} 
\sO_{bc} e^{- i (t_1 - t_2) E_c} \sO_{cd} e^{- E_d ( i t_2 +  \b_R/2)} \OO_{da} \\
&=& {e^{-N^2 f_0}  \ov Z} 
\sum_{a,b,c,d} e^{- \b_L  E_a - \ha \b_R (E_b +E_d) }  e^{- i (t_1 - t_2) E_c  + i t_1 E_b - i t_2 E_d}
e^{-i N^2 (g_{ba} - g_{da})}  \sO_{bc}  \sO_{cd} ,
\label{onex}
\eea
where $\sO_{ab} = \vev{a|\sO|b}$. 

Let us now suppose that, in the large $N$ limit,  the rapid oscillations of the phase suppress all other terms, so that the dominant contribution to the sum in~\eqref{onex} arises  from the sector in which the phase vanishes,\footnote{This is in fact a subtle point. A detailed discussion is deferred to Sec.~\ref{sec:largeN} to avoid breaking the conceptual flow here.} namely from the terms with 
$d=b$. We then find  
\bea 
G_{RR} &= & {e^{-N^2 f_0}  \ov Z}  \sum_{a, b,c} e^{- \b_L  E_a } e^{- \b_R  E_b} 
\sO_{bc} e^{- i (t_1 - t_2) (E_c- E_b)} \sO_{cb} \\
&=&   {1 \ov Z_{\b_R}} \Tr \le(e^{-\b_R H} \sO(t_1) \sO(t_2) \ri) 
 \label{ehn1}
\eea  
Similarly, we find in the large $N$ limit,  
\bea
\vev{\Psi_\OO|\sO_L (t_1) \sO_L (t_2)|\Psi_\OO} &=&  {1 \ov Z_{\b_L}} \Tr \le(e^{-\b_L H} \sO(t_1) \sO(t_2) \ri), \\
 \vev{\Psi_\OO|\sO_L (t_1) \sO_R ( t_2)|\Psi_\OO}  &=&  \vev{\sO}_{\b_R} \vev{\sO}_{\b_L}  \ .
 \label{ehn3}
\eea
From equations~\eqref{ehn1}--\eqref{ehn3}, $\sY_R$ and $\sY_L$ are in thermal states with inverse temperatures $\b_R$ and $\b_L$, respectively, with there being no correlation between $\sY_R$ and $\sY_L$. Clearly, $\om_{\Psi_\OO}$ is a mixed state for $\sY = \sY_R \otimes \sY_L$, which means that it has a nontrivial commutant. 

From~\eqref{ehn1}--\eqref{ehn3}, the commutant $\sY'$ should have the structure $\sY' = \sY_R' \otimes \sY_L'$ where $\sY_R'/\sY_L'$ is used to purify $\sY_R/\sY_L$ in a thermal state with inverse temperature $\b_R/\b_L$. This agrees well with the gravity 
description (recall Sec.~\ref{sec:rev} and  Fig.~\ref{fig:baby}(a)). 

We can also calculate the entanglement entropies. For example, for the second R\'enyi entropy $S_2^{(R)}$ of CFT$_R$, we find 
\ie 
 e^{- S_2^{(R)}}
& = \Tr \rho_R^2 = {1 \ov Z^2} \Tr  \le(e^{-\b_R H} \OO e^{- \b_L H}  \OO^\da e^{- \b_R H}  \OO e^{- \b_L H}  \OO^\da \ri) \cr
& = {1 \ov Z_{\b_L}^2 Z_{\b_R}^2} \sum_{a,b,c,d}  e^{-\b_R (E_a +E_c) - \b_L (E_b + E_d)}  e^{i N^2 (g_{ab} - g_{cb} + g_{cd} - g_{ad})} \cr
& = {1 \ov Z_{\b_L}^2 Z_{\b_R}^2} \sum_{a,b,c,d}  e^{-\b_R (E_a +E_c) - \b_L (E_b + E_d)}  
(\de_{ac} + \de_{bd}) \\
& = e^{-S_{2, \b_R}} + e^{-S_{2, \b_L}} 
\label{Mrenyi}
\fe 
which again agrees with the gravity result~\eqref{renyiE}.\footnote{As~\eqref{renyiE}, the calculation~\eqref{Mrenyi} does not apply at extremely low temperatures. See Sec.~\ref{sec:avergN} for further discussion.}


Similar calculations of correlation functions can also be performed using the more general ansatz~\eqref{newsan}, which captures some  qualitative features of the gravity description for arbitrary 
$\mu_\OO$; see Appendix~\ref{app:generalization} for details. 

To summarize, the truncation~\eqref{ps1} is compatible with the existence of a baby universe, and captures the essential features of the gravity description, provided that $N$-dependent phases are allowed in the coefficients.

It is worth unpacking a bit further the ``magic'' of such $N$-dependent oscillatory behavior, i.e., where the baby universe comes from. 
 The action of $\OO$ on a low-energy state, when projected back to the low-energy sector,
\be\label{basec}
\OO \ket{a} = \sum_b \OO_{ba} \ket{b} \sim \sum_b e^{i N^2 g_{ba}} \ket{b},
\ee
does not in fact generate states with a well-defined large-$N$ limit. Despite appearances, these are not genuine low-energy states of the vacuum-sector Hilbert space, due to $N$-dependence.  Rather, $\OO \ket{a}$ should be regarded as belonging to a new sector of states that themselves lack a well-defined large-$N$ limit. It is precisely these new sectors (for both $R$ and $L$) that ``generate'' the Hilbert space of the baby universe. In other words, the baby universe can be viewed as the low-energy manifestation of such sectors.

Alternatively, instead of~\eqref{newsan}, we can consider the ansatz~\cite{Sas22,AntSas23,AntRat25} 
\be \label{oldsan}
\OO_{ab} = e^{-\ha N^2 f_0} f(E_a, E_b) \sR_{ab},
\ee
where $\sR_{ab}$ is a Gaussian random matrix with average 
\be \label{ensc}
\overline{\sR_{ab}^*  \sR_{cd}} = \de_{ac} \de_{bd} \ . 
\ee
We would obtain the same results as~\eqref{ehn1}--\eqref{Mrenyi} by starting from~\eqref{oldsan} and performing ensemble averages in the evaluation of these correlators. Although the mathematical expressions coincide, the conceptual interpretations are different. In~\eqref{ehn1}--\eqref{Mrenyi} there is no ensemble average over different states or theories. However, as will be discussed in detail in Sec.~\ref{sec:largeN}, the large $N$ limit is subtle and involves some averages. 
 Mathematically, the effects of highly fluctuating phases can be approximated by ensemble averages in the large-$N$ limit, but this should be understood as a computational device rather than as an averaging over different theories.

To conclude this subsection, we emphasize that equation~\eqref{newsan} is at this stage a postulated approximation, supported only by the agreement of its implications with gravitational results. Ultimately, it should be substantiated or refuted through an analysis of 
the matrix elements of a general heavy operator  in a CFT.


\subsection{The nature of the large $N$ limit} \label{sec:largeN}

A central assumption in Sec.~\ref{sec:comm} is that the rapidly oscillating phases appearing in~\eqref{onex} and~\eqref{Mrenyi} effectively suppress non-stationary-phase contributions in the 
large $N$ limit. We now turn to a more detailed examination of this assumption.\footnote{I would like to thanks Elliott Gesteau for  discussion on this part.} 

Consider a sum of the form 
\be \label{fn}
F(N) = \sum_a f_a  e^{i N^2 g_a}, \quad f_a > 0, \quad \sum_a f_a < \infty ,
\ee
which serves as a model for the sums appearing in~\eqref{onex} and~\eqref{Mrenyi}. 
The usual intuition is that the non-stationary-phase contributions to~\eqref{fn} cancel in the limit, leading to
\be \label{fnl}
\lim_{N \to \infty} F(N) = \sum_{a \in I} f_a, \qquad I = \{ a | g_a = 0 \},
\ee
which is the result used in Sec.~\ref{sec:comm}.

However, \eqref{fnl} is incorrect for a {\it pointwise} $N \to \infty$ limit; in fact, such a limit does not exist. For each fixed $N$, the sum \eqref{fn} is absolutely convergent, and because $N$ appears {\it only} in the phases, $F(N)$ is a highly oscillatory (typically quasi-periodic) sequence in $N$, and thus admits no pointwise $N \to \infty$ limit. Consequently, the sums in \eqref{onex} and \eqref{Mrenyi} likewise do not have a {\it pointwise} $N \to \infty$ limit.

Relatedly, Gesteau proved a theorem~\cite{Ges25} asserting that if the correlation functions of single-trace operators in the state $\ket{\Psi_\OO}$ admit a well-defined pointwise $N \to \infty$ limit, then $\ket{\Psi_\OO}$ converges to \eqref{ps1} with $N$-independent coefficients $\psi_{ab}$. In particular, in the $N \to \infty$ limit, the state $\ket{\Psi_\OO}$ becomes pure in $\sH_\Om^{(R)} \otimes \sH_\Om^{(L)}$, which would preclude a semiclassical description of the baby universe from the boundary.

The limit~\eqref{fnl} can still be made sense of if we replace the pointwise $N \to \infty$ limit with an averaged one. More explicitly, consider, for example,  
\be
\widetilde F(N) = \frac{1}{W_N} \sum_{k=0}^{W_N} F(N+k) ,
\ee
where $W_N$ denotes the window size for average which should diverge with $N$, though it can grow arbitrarily slowly, e.g. as $\log \log N$. With such an averaging prescription, the oscillatory part of $F(N)$ is smoothed out, and $\widetilde F(N)$ admits a well-defined $N \to \infty$ limit. For convenience, we introduce a new notation for the averaged large-$N$ limit:
\be \label{fnlN}
\widetilde{\lim_{N \to \infty}} F(N)  \equiv \lim_{N \to \infty} \frac{1}{W_N} \sum_{k=0}^{W_N} F(N+k), 
\ee
which yields 
\be
\widetilde{\lim_{N \to \infty}} F(N) = \sum_{a \in I} f_a,
\qquad I = \{ a \mid g_a = 0 \} \ .
\ee
Unlike the pointwise limit, which does not exist, this averaged limit is well defined and isolates precisely the stationary-phase contributions.

We propose that such an averaged large-$N$ limit is the appropriate one for comparison with the gravity description of the baby universe. Physically, this appears quite natural: semi-classical gravity captures the large-$N$ behavior of the boundary theory, but what it reflects is the universal behavior valid for all sufficiently large $N$, with erratic features tied to specific values of $N$ effectively filtered out. 

Large $N$-averages were advocated previously in~\cite{SchWit22} as an alternative to ensemble averages over different boundary theories, in the context of processes involving black hole states.
The proposal of~\cite{SchWit22} for bulk Euclidean path integrals with multiple boundaries can be expressed as
\be \label{sgan}
\wt{\lim_{N \to \infty}} Z_N (\beta_1, R_1) \cdots Z_N (\beta_n, R_n) = \sum \text{wormhole contributions} ,
\ee
where $Z_N(\beta, R)$ denotes the boundary partition function at finite $N$, with (complex) inverse temperature $\beta$ and additional quantum numbers collectively denoted by $R$. In~\eqref{sgan}, $\wt{\lim_{N \to \infty}}$ should be understood as some averaged 
large-$N$ limit, not necessarily of the exact form of~\eqref{fnlN}. Importantly, the left-hand side is not factorized, since the averaging over $N$ has already been performed. This non-factorization provides the boundary counterpart of wormhole contributions in the bulk.



With an averaged large-$N$ prescription in mind, the ansatz of~\eqref{newsan} can in fact be significantly generalized. The phase postulated in~\eqref{newsan} is only one specific example of $N$-dependent oscillatory behavior, and it need not be restricted to phases. The essential point is that $\OO_{ab}$ can, in principle, exhibit $N$-dependent oscillations, so that it, and consequently the corresponding correlation functions, do not possess a pointwise large-$N$ limit.

It is instructive to compare PETS with the more familiar TFD state. Below the Hawking-Page temperature, the TFD state defined in~\eqref{tfd} has a well-defined large-$N$ limit,\footnote{In this regime the dominant contributions come from states with energies of order $O(N^0)$, so the state truncates to this sector in the large-$N$ limit.} as does the state $\omega_{\Psi_\beta}$ defined in~\eqref{Dsta} through correlation functions.

Above the Hawking-Page temperature, the TFD state itself does not admit a large-$N$ limit, since contributions from states with energies of order $O(N^2)$ dominate. Yet $\omega_{\Psi_\beta}$ still has a pointwise large-$N$ limit, with the collective effects of $O(N^2)$ states giving rise to the horizon, the black hole interior, and spacetime connectivity~\cite{LeuLiu21a,LeuLiu21b,EngLiu23}.

For PETS, the situation is different. From~\eqref{basec}, neither the original state~\eqref{PETS} nor the state $\omega_{\Psi_\OO}$ defined through correlation functions admits a pointwise large-$N$ limit. Nevertheless, $\omega_{\Psi_\OO}$ does have an averaged large-$N$ limit, in which the effects of the heavy operator are encoded collectively, giving rise to the baby universe.

From now on, when referring to the large $N$ limit, we will have in mind an averaged limit like~\eqref{fnl}.


\subsection{Finite $N$ origin of operators in the baby universe} 

In the Sec.~\ref{sec:comm} we gave a boundary argument demonstrating the existence of operators beyond those generated by single-trace operators in the large-$N$ limit. These additional operators correspond to physical operations in the baby universe. 
Heuristically, they can be understood as the operators acting on the sector~\eqref{basec}. 
In this subsection, we discuss their possible finite-$N$ origin in more rigorous algebraic terms. 
  
At finite $N$, the algebras for the $R$ and $L$ CFTs are given by $\sM_R \equiv \sB (\sH_R)$ and $\sM_L \equiv \sB(\sH_L)$, with $\sM_L = \sM_R'$.  The TFD state $\ket{\Psi_\b}$ is cyclic and separating with respect to $\sB (\sH_R)$ and $\sB(\sH_L)$.  
The corresponding modular operator is
\be
\De_\b = e^{-\b (H_R - H_L)}, 
\ee
where $H_R, H_L$ are the Hamiltonians of the $R$ and $L$ CFTs, 
and the modular conjugation operator $J_\b$ acts as 
\be 
J_\b A_R J_\b = \tilde A_L , \quad A \in \sB (\sH), \quad \tilde A = \Th A \Th  \ .
\ee

Now suppose that $\OO$ is such that $\ket{\Psi_\OO}$ is cyclic and separating with respect to both $\sB(\sH_R)$ and $\sB(\sH_L)$.\footnote{This is the case, for example, if $\OO$ is invertible.} We denote the corresponding modular operator and modular conjugation by $\Delta_\OO$ and $J_\OO$, respectively. In the large-$N$ limit, $\ket{\Psi_\OO}$ corresponds to the state $\ket{\mathbf{1}}_{\Psi_\OO}$ (arising from the identity operator in the GNS construction) in the GNS Hilbert space, $\sH_{\Psi_\OO}^{\text{(GNS)}} = \sH_{\Psi_\OO}^{\text{(Fock)}}$.\footnote{$\ket{\mathbf{1}}_{\Psi_\OO}$ plays the role of the ``vacuum'' in the bulk Fock space.} We expect that $\ket{\mathbf{1}}_{\Psi_\OO}$ is likewise cyclic and separating with respect to $\sX_R$ and $\sX_L$. We will denote by $\tilde \Delta_\OO$ the modular operator associated with the pair $(\ket{\mathbf{1}}_{\Psi_\OO}, \sX_R)$.

Consider the action of modular flow generated by $\De_\OO$ on a single-trace operator 
\be 
\De_\OO^{i s} \sO_R \De_\OO^{- i s}, \quad s \in \RR \ .
\ee
It has been conjectured in~\cite{LeuLiu22} that for a general semi-classical state, such modular-flowed operators also have a well-defined large-$N$ limit in the state,\footnote{Further support will be provided in upcoming work with Justin Berman.} and furthermore, 
\be
\lim_{N \to \infty} \pi_{\Psi_\OO} \le(\De_\OO^{i s} \sO_R \De_\OO^{- i s} \ri) 
= \tilde \De_{\OO}^{is} \pi_{\Psi_\OO}  (\sO_R) \tilde \De_{\OO}^{-is}  \ .
\ee

In the long-BH phase of Fig.~\ref{fig:baby}(b), $\sY_R$ is dual to the region outside the right horizon. In this case,  $\ket{\bid}_{\Psi_\OO}$ is cyclic and separating with respect to $\sY_R$ as well. A theorem in von Neumann algebra then warrants that\footnote{T. Faulkner, private communication.} 
\be \label{scr1}
\sX_R = \{\tilde \De_{\OO}^{is} \sY_R \tilde \De_{\OO}^{-is}, \; s \in \RR\} \ .
\ee 
This implies that 
\be \label{scr2}
\De_\OO^{i s} \sS_R \De_\OO^{- i s}, \quad s \in \RR ,
\ee
can generate all the operators in $\sA_R$ not in $\sS_R$. Similarly with $\sA_L$. Thus all the operators in the interior of the horizons should descend from such modular flowed operators. 

In the TAdS-BU phase, $\sY_R = \sB (\sH_\Om^{(R)})$ is type I, and from~\eqref{ehn1}--\eqref{ehn3}, its action on $\ket{\bid}_{\Psi_\OO}$ cannot generate the states in $\sH_\Om^{(L)}$. Thus, $\ket{\bid}_{\Psi_\OO}$ is not cyclic with respect to $\sY_R$.
In this case, although there is no theorem guaranteeing~\eqref{scr1}, it can be generally expected that modular-flowed operators~\eqref{scr2} might nevertheless generate all operators in $\sA_R$.\footnote{In fact, one expects that any generic one-parameter automorphism of $\sA_R$, including modular flow, will be sufficient to generate $\sA_R$ from $\sS_R$ (E. Witten, private communication). An alternative argument can also be made based on the existence of a Euclidean description of the state.}




In the case of PETS, the modular operator $\De_\OO$ 
can be obtained formally using $\OO$. More explicitly, we have  
\bega\label{rhor}
\rho_R = {1 \ov Z} e^{-\ha \b_R H_R} \OO_R e^{- \b_L H_R}  \OO_R^\da e^{-\ha \b_R H_R} \\
\rho_L = {1 \ov Z} e^{-\ha \b_L H_L} \tilde \OO_L^\da e^{- \b_R H_L}  \tilde \OO_L  e^{-\ha \b_L H_L}  , \quad
\tilde \OO = \Th \OO \Th, 
\\
\De = \rho_R \rho_L^{-1} =  e^{-\ha \b_R H_R} \OO_R e^{- \b_L H_R}  \OO_R^\da e^{-\ha \b_R H_R} 
e^{\ha \b_L H_L} \tilde \OO_L^{-1} e^{ \b_R H_L}  \tilde \OO_L^{-1 \da}  e^{\ha \b_L H_L} ,
\end{gather} 
and 
\be \label{egb}
\De_\OO^{is} \sO_R \De_\OO^{-is} = \rho_R^{is} \sO_R \rho_R^{-is} \ .
\ee
While the modular operator is complicated, in essence what the flow~\eqref{egb} does is to use $\OO$ to ``dress'' $\sO_R$ 
in a certain way,\footnote{Note that various factors involving exponentials of $H_R$ in~\eqref{rhor} only generate ordinary time translations.} which is consistent with the heuristic picture that such operators  act on the sector~\eqref{basec}. 

In the special case $\b_R = {\b \ov 2}$, where the insertion of $\OO (i \b_R/2)$ is in at middle
point of the Euclidean half-circle, $\ket{\Psi_\OO}$ is invariant under the modular conjugation of the TFD state, 
\be 
J_\b \ket{\Psi_\OO} =  \ket{\Psi_\OO} \ .
\ee
In this case\footnote{$\ket{\Psi_\OO}$ lies in the natural cone of the pair $(\sB (\sH_R), \ket{\Psi_\b})$.}
\be 
J_\OO = J_\b, 
\ee
which acts on the Euclidean time circle by reflection with respect to the central vertical axis.

In~\cite{AntSas23}, it was pointed that Euclidean operator insertions such as 
\be \label{ejj} 
\sO (i \tau) \OO(i \b_R/2) \ket{\Psi_\b} \; (\tau < \b_R/2) , \qquad 
\OO(i \b_R/2) \sO (i \tau)  \ket{\Psi_\b} \; (\tau > \b_R/2)  ,
\ee
can be used to generate bulk states with excitations in the baby universe. 
However, we emphasize that Euclidean operators such as $\sO(i \tau)$ cannot be used to reconstruct bulk operators, since they are not well-defined operators. As seen from~\eqref{ejj}, in order to obtain a well-defined state, operators in Euclidean time must be {\it time-ordered}. If $\sO(i \tau)$ were a well-defined operator, then $\sO (i \tau) \ket{\Psi_\OO} = \sO (i \tau) \OO(i \b_R/2) \ket{\Psi_\b}$  would be a well-defined vector for arbitrary $\tau$, which is not the case.

\subsection{Absence of a geometric entanglement wedge and failure of QES at $O(N^0)$} \label{sec:AEW}


In this subsection, we argue that in the TAdS-BU phase, in contrast to the long-BH phase, the $R$ or $L$ boundary does not have a {\it geometric} entanglement wedge. In other words, there does not exist a geometric region in the bulk whose physics is equivalent to that of the $R$ or $L$ boundary in the large $N$ limit. 


Consider the entanglement entropy $S_R$ for CFT$_R$ at the time-reflection symmetric slice $t=0$, which can be calculated on the gravity side using the replica trick~\cite{AntSas23}. In contrast to the situations discussed in~\cite{LewMal13,FauLew13}, in this case there is no codimension-two invariant submanifold under the replica symmetry in the bulk (and thus the entropies are $O(N^0)$). Moreover, there is no bulk subregion whose entanglement entropy can be identified with that of the $R$-boundary.
Thus there is no entanglement wedge.\footnote{Note that this is different from the situation that the quantum extremal surface exists but is an empty set.} 


 
The same conclusion can also be reached by applying the quantum extremal surface~(QES) prescription~\cite{EngWal14} to 
compute $S_R$. It is simplest to illustrate this in the $\mu_\OO \to \infty$ limit, where AdS$_R$ and AdS$_L$ are in thermal states with inverse temperatures $\beta_R$ and $\beta_L$. Possible choices of QES and the entanglement wedge $\sW_R$ are: 
(1) There is a nontrivial QES in the baby universe, with $\sW_R = \text{AdS}_R \cup \Ga_R$ where $\Ga_R$ is a subregion in the baby universe. In this case, we have $S_R = O(N^2)$ as the QES would contribute a nonzero area term. 
 (2) An empty QES with $\sW_R = \text{AdS}_R$. In this case, we have $S_R = S_{\b_R}$, which is the entropy of the AdS$_R$. 
  (3) An empty QES with $\sW_R = \text{AdS}_R \cup BU$, where $BU$ denotes the full baby universe. In this case, we have 
  $S_R = S_{\b_L}$, which is the entropy of the AdS$_L$, from that the full state is a pure state. (1) would never be a minimal QES and we thus have $S_R =  {\rm min} (S_{\b_R}, S_{\b_L})$, which does not agree with~\eqref{renyiE}. We thus conclude that the QES prescription fails in this case. 
  
We expect that the absence of an entanglement wedge for the $R$ or $L$ system is generic in situations with $O(N^0)$ entanglement entropy and a baby universe. More specifically, one can argue that the QES prescription is generally valid only at order $O(N^2)$ and should be applied with caution when analyzing cases with $O(N^0)$ entanglement entropy.\footnote{Work to appear.}
The absence of entanglement wedge for the $R$ and $L$ boundaries also invalidates an argument for the one-dimensional Hilbert  space for the closed universes, on which we will comment further in Sec.~\ref{sec:one-dim}.

We now further strengthen the case by arguing that a bulk local operator in the baby universe must be expressed in terms of operators of both $R$ and $L$ CFTs. 


For this purpose, we first examine the structure of boundary algebra $\sX = \sX_R \lor \sX_L$ more closely. 
At finite $N$, the full operator algebra is given by $\sB (\sH_R) \otimes \sB (\sH_L)$ with both factors being type I. 
In the large $N$ limit, 
given that the entanglement and R\'enyi entropies between CFT$_R$ and CFT$_L$ are $O(N^0)$---i.e., remain finite as $N \to \infty$, the algebras $\sX_R$ and $\sX_L$  should be type I. The existence of well-defined entropies (without the ambiguity of adding arbitrary constants) suggests the existence of a well-defined trace, with no ambiguity in its definition.
 
 This implies that the GNS Hilbert space $ \sH_{\Psi_\OO}^{(\rm GNS)} $ can be factorized, i.e., 
 there exist Hilbert spaces $\wt \sH_R$ and $\wt \sH_L$ such that 
\be \label{heg2}
 \sH_{\Psi_\OO}^{(\rm GNS)} = \wt \sH_R \otimes \wt \sH_L, \quad
 \sX_R = \sB (\wt \sH_R ), \quad \sX_L = \sB ( \wt \sH_L) \ .
 \ee
 Combining~\eqref{heg2} with~\eqref{heg}--\eqref{heg1}, we conclude that the baby universe Hilbert space can be further factorized. 
 That is, there should exist Hilbert spaces 
 $\sK_R$ and $\sK_L$ such that
 \bega \label{fact1}
 \sH_{\rm BU}^{(\rm Fock)} = \sK_R \otimes \sK_L, \quad
 \wt \sH_R = \sK_R \otimes \sH_\Om^{(R)} , \quad \wt \sH_L = \sK_L \otimes \sH_\Om^{(L)} , \\
 \sX_R = \sB (\sK_R) \otimes \sY_R, \quad \sX_L = \sB (\sK_L) \otimes \sY_L , \quad 
 J_\OO \sX_R J_\OO = \sX_L \ .
 \end{gather} 
 
We denote the right and left halves of the baby universe as $BU_R$ and $BU_L$, respectively. At first sight, one might be tempted to identify the bulk operator algebra $\sM_{BU_R}$ in $BU_R$ with $\sB(\sK_R)$ (and similarly for $L$). However, this cannot be correct: $\sM_{BU_R}$ is of type III$_1$ (as $BU_R$ is a subregion), whereas $\sB(\sK_R)$ is of type I. The same reasoning applies to any subregion of $BU_R$. Consequently, bulk operators in $BU_R$ ($BU_L$) must necessarily involve operators from CFT$_L$ (CFT$_R$). This confirms that there is no geometric entanglement wedge associated with the $R$ and $L$ boundaries.


Equation~\eqref{fact1} implies that the Hilbert space of the baby universe can be further factorized. This can indeed be seen from the gravity side.  Consider $t=0$ slice of the baby universe in Fig.~\ref{fig:baby}(a) which is obtained by 
patching together two copies of hyperbolic space 
\be \label{einn}
ds^2_H = d \xi^2 + \sinh^2 \xi d \Om_{d-1}^2 , \quad \xi < \xi_0
\ee
along $\xi = \xi_0$. 
The full space is topologically a $d$-dimensional sphere and is compact. 
Now consider a scalar field $\phi$ in the baby universe. We denote $\phi$ in two copies of~\eqref{einn} respectively as 
$\phi_r$ and $\phi_l$, which satisfy the continuity condition at $\xi = \xi_0$, 
\be 
\phi_r (\xi_0) = \phi_l (\xi_0) \ .
\ee
From $\phi_{r, l}$ we can construct 
\bega 
\phi_s = \phi_r (\xi) + \phi_l (\xi), \quad 
\phi_a = \phi_r (\xi) - \phi_l (\xi) 
\end{gather} 
which satisfy, respectively, the free boundary condition and the Dirichlet boundary condition $\phi_a(\xi_0) = 0$ at $\xi = \xi_0$.
$\phi_s$ and $\phi_a$ can be quantized independently to generate Hilbert spaces $\wt \sK_s$ and $\wt \sK_a$, i.e., 
\be 
\sH_{\rm BU}^{(\rm Fock)} = \wt \sK_s \otimes \wt \sK_a \ .
\ee
Note that we cannot directly identify $\wt \sK_{s,a}$ with $\sK_{R, L}$ as the oscillators for them could in principle be related by a Bogoliubov transformation. 

In the $\mu_\OO \to \infty$ limit, $\xi_0 \to \infty$, the $t=0$ slice of the baby universe essentially consists of two copies of global AdS. However, there is a crucial difference between boundary algebras $\sB (\sK_R)$ and $\sB (\sK_L)$ describing the baby universe from $\sY_R$ and $\sY_L$. For any finite $\xi_0$,\footnote{Physically $\mu_\OO$ is always finite.} no matter how large, the IR part of global AdS with $\xi > \xi_0$ is excised. This means that the boundary algebra describing the baby universe does not have any ``local'' boundary operator, i.e., it has better ``UV behavior'' than $\sY_R$ or $\sY_L$.

\subsection{Comments on large finite $N$}

Our discussion above has focused on the $N \to \infty$ limit. We now briefly comment on the case of large but finite $N$, corresponding to small but finite $G_N$.

Consider first the TFD state. At large but finite $N$, the TFD state at a given temperature includes contributions from both the low-energy thermal AdS sector and the high-energy black hole sector. Below the Hawking-Page temperature, the system is well approximated by thermal AdS, though with an exponentially small probability of transitioning into a black hole. In this regime, the states in $\sH_\Omega^{(R)} \otimes \sH_\Omega^{(L)}$ and the associated algebra $\sY = \sB (\sH_\Omega^{(R)} ) \otimes \sB(\sH_\Omega^{(L)})$ provide a good description of low-energy excitations, subject to finite-$N$ corrections.

A similar picture applies for PETS. At finite $N$, both the TAdS-BU sector and the long-BH sector are part of the full Hilbert space, again with finite-$N$ corrections. At low temperatures, the low-energy excitations can be approximately described by $\sH_{\Psi_\OO}^{(\rm GNS)}$ and the associated algebra $\sX = \sX_R \otimes \sX_L$, with  an exponentially suppressed probability of transitioning into the long-BH phase. In particular, the baby universe should continue to admit a semiclassical Hilbert space.

\section{Comments on the Maldacena-Maoz Closed Universe} \label{sec:MM}

In this section we make some comments on the closed universes constructed in~\cite{MalMao04} showing that there is a natural holographic description in terms of certain algebra and state.

\subsection{Review of the setup} 

\begin{figure}
\begin{center}
\includegraphics[width=14cm]{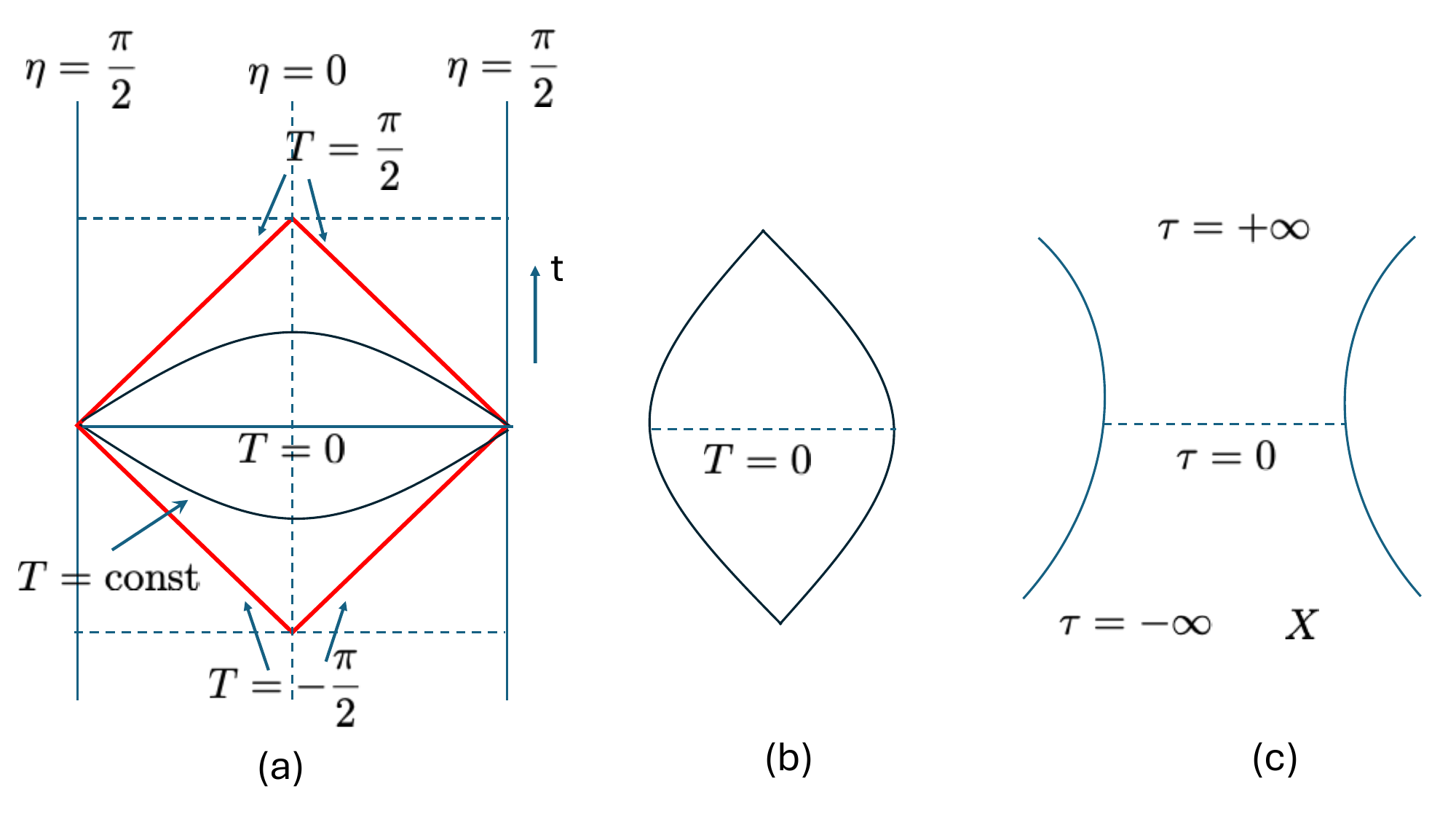}
\caption{\small (a) The part of AdS covered by the 
coordinates~\eqref{frws} lies inside the red diamond in the plot. Each point in figure represents a
$S_{d-1}$. Various constant $T$ surfaces are show in the figure.  Note
that $T=0$ hypersurface coincides with $t={0}$ surface. (b) The closed universe~\eqref{euw} resulted from the quotient. (c) The corresponding Euclidean manifold~\eqref{euw1} resulted from the quotient. Inserting boundary operators (e.g. $X$) at $\tau =-\infty$ can create  ``excited'' states in the Fock space of closed universe. 
}
\label{fig:AdSFRW}
\end{center}
\end{figure}

Consider global AdS$_{d+1}$ 
\be \label{Gads}
{ds^2 } =  {1 \ov \cos^2 \eta} \le(-dt^2 + d \eta^2 + \sin^2 \eta d \Om_{d-1}^2 \ri),
\ee
which can be written 
using an open FRW slicing as  
 \be\label{frws}
 {ds^2} = -dT^2 + \cos^2 T \, dH_{d}^2 = -dT^2 + \cos^2 T
 (d\rho^2 + \sinh^2 \rho d \Om_{d-1}^2),
 \ee
where $dH_d^2 $ denotes the metric of $d$-dimensional hyperbolic space $\HH_d$ and  $T \in (-{\pi \ov 2}, {\pi \ov 2})$,  
$\rho \in (0, \infty)$. This coordinate system covers only a single Wheeler-DeWitt (WDW) patch (as illustrated in Fig.~\ref{fig:AdSFRW}(a));  the spatial hyperbolic space shrinks to zero size at $T= \pm {\pi \ov 2}$, but these are merely coordinate singularities. 
The WDW patch intersects with the boundary at a single boundary time slice $S_{d-1}$ at $T =0$ and $\eta \to +\infty$. 

We can analytically continue~\eqref{frws} to the Euclidean signature by 
taking $T \to - i \tau$, with $\tau \in (-\infty, +\infty)$, which results in a $d+1$-dimensional hyperbolic space $\HH_{d+1}$ 
\be \label{eucH}
{ds^2_E } = d \tau^2 + \cosh^2 \tau \, dH_{d}^2 = d\tau ^2 + \cosh^2 \tau
 (d\rho^2 + \sinh^2 \rho d \Om_{d-1}^2) \ .
 \ee
The metric~\eqref{eucH} covers the full hyperbolic space, with the boundary topologically $S_d$, consisting of 
two copies of $\HH_{d}$ lying at $\tau \to \pm \infty$, patched together at the $S_{d-1}$ of $
\tau =0, \rho \to \infty$,

The MM construction quotients a constant-$T$ hyperbolic space $\HH_d$ by a free-acting discrete group $\Ga$
such that the resulting quotient manifold $M_d = \HH_d/\Ga$ is {\it compact}, resulting in
\bea\label{euw}
{ds^2 } &=&  -dT^2 + \cos^2 T \, dM_{d}^2 , \\ 
{ds^2_E} &= & d\tau^2 + \cosh^2 \tau \, dM_{d}^2  \ .
\label{euw1}
\eea
In the Lorentzian description (see Fig.~\ref{fig:AdSFRW}(b)), we find a closed universe with no boundary, and $T = \pm {\pi \ov 2}$ become genuine big bang and big crunch curvature singularities. In the Euclidean signature (see Fig.~\ref{fig:AdSFRW}(c)), the boundary consists of two copies of $M_d$, lying at $\tau \to \pm \infty$, which are now disconnected.

\subsection{Holographic description}

To give a holographic description of the closed universe constructed above, we start with the boundary description of~\eqref{frws} before performing the quotient. Since the WDW patch covers a bulk Cauchy slice, the bulk system in the patch in the $G_N \to 0$ limit is described by the full boundary algebra $\sM \equiv \sB (\sH_\Om)$, where $\sH_\Om$ is the boundary GNS Hilbert space in the vacuum sector (which is identified with the bulk Fock space for global AdS). The boundary algebra $\sB (\sH_\Om)$ is in turn generated by a generalized free field theory of single-trace operators.\footnote{Perturbative $1/N$ corrections can also be incorporated by taking into account of high order OPEs among single-trace operators.} 

The bulk isometry group $G$, which is identified as the boundary conformal group, acts 
on $\sM$ by unitary automorphisms. That is, for an element $g \in G$, there exists a unitary operator $U_g \in \sM$ such that $g \cdot A = U_g A U_g^\da$ for $A \in \sM$.  Performing the quotient by $\Ga \subset G$ in the bulk can be realized in the boundary theory by performing the quotient on the generalized free field theory.\footnote{We thank J.~Maldacena for discussions.} 
More explicitly, for a single-trace operator $\sO$, we can construct its invariant version $\hat \sO$ by summing over the images under the action of $U_g$. We define an invariant algebra $\sM_\Ga$ 
as 
\be \label{invA}
\sM_\Ga = \text{the algebra generated by all the} \; \hat \sO  \ .
\ee
The state on $\sM_\Ga$ for the closed universe descends from the state of the parent theory, i.e., the vacuum state $\ket{\Om}$, we define 
\be \label{gsta}
\om_\Ga (A) = \vev{\Om|A|\Om}, \quad A \in \sM_\Ga  \ .
\ee
The GNS Hilbert space $\sH_\om^{(\rm GNS)}$ resulting from the above action can be identified with the 
Hilbert space of the closed universe in the $G_N \to 0$ limit. Bulk reconstruction of a field operator in the closed universe in terms of boundary operators should descend from that of the un-quotiented theory.

It will be instructive to consider an explicit example of $\Ga$ and work out the algebra $\sM_\Ga$ explicitly. 
We will leave an explicit construction for future work.  

Note that in terms of the global coordinates~\eqref{Gads}, an element of $\Ga$ acts on the global time $t$.  
Thus, an element of $A$ of the invariant algebra~\eqref{invA} should contain certain kind of smearing in the time direction. In particular,  local operators dual to a bulk operator near the AdS boundary are projected out by the quotient.  
We thus expect the resulting invariant algebra has better ``UV'' behavior than the original algebra. It would certainly be desirable to understand this notion more precisely in an explicit example.

Now consider the case of large but finite $G_N$. In this setting, the algebra for the parent theory is given by 
$\sM = \sB(\sH)$, where $\sH$ denotes the full CFT Hilbert space, so that $\sM$ is the complete operator algebra of the boundary system. It is not clear how to perform the quotient on $\sM$. Nevertheless, irrespective of the precise procedure, there should still exist a sector describing low-energy excitations above the ``vacuum''---namely, the states in $\sH_\omega^{(\mathrm{GNS})}$---though now subject to finite-$N$ corrections.



\subsection{Connection with Euclidean path integrals} \label{sec:connE}

An alternative way to construct  states in the bulk Fock space $ \sH_\omega^{(\mathrm{GNS})}$ for the closed universe  is via a Euclidean path integral with boundary operators inserted on the Euclidean boundary at $\tau = -\infty$. See Fig.~\ref{fig:AdSFRW}(c). More explicitly, denoting the state obtained by inserting operator $X$~(which could be a product of local operators) at $\tau =-\infty$ as $\ket{X}$, we have 
\be  \label{overl}
\vev{X'|X} = A (X', X) \equiv \int_{X}^{X'}   D \Phi (\tau) \,  e^{- S_E [\Phi]}  , \qquad \ket{X}, \ket{X'} \in \sH_\omega^{(\mathrm{GNS})},
\ee
where $\Phi$ collectively denotes bulk fields defined on the  Euclidean geometry in~\eqref{euw}, with $S_E$ their corresponding action. 
We stress that~\eqref{overl} is the path integral for the bulk quantum field theory on the Euclidean background of~\eqref{euw}, not the full gravitational path integral.  The boundary interpretation of the right hand side of~\eqref{overl} 
has been a long-standing puzzle~\cite{WitYau99,MalMao04}, as it involves ``correlations'' between observables in different Euclidean CFTs. 


To put the right hand side of~\eqref{overl} in proper context, consider the following quantity 
\be \label{apar}
F (X', X) \equiv  \int_{M_d, X}^{M_d, X'} D g D \Phi \, e^{- S_E [g, \Phi]} 
\ee
which is defined by formally integrating over all $(d+1)$-dimensional Euclidean manifolds with two copies of $M_d$ as its boundaries (and boundary conditions specified by $X$ and $X'$ insertions). Here one also integrates over all metrics $g$. 
$M (X', X)$ is related to the fluctuation part of one of the saddle contribution to $F(X', X)$ in the large $N$ limit.  
More explicitly, in the saddle point approximation to~\eqref{apar}, 
\be \label{ove1}
F (X', X) = \cdots + e^{- S_E [K]} A (X', X) + \cdots ,
\ee 
where $K$ denotes the $(d+1)$-dimensional Euclidean spacetime of~\eqref{euw}, and $\cdots$ denote contributions from other saddles, including disconnected geometries. 
The saddle $K$ may not be the dominant contribution. For example, as discussed in~\cite{SchWit22}, in the case $d=2$ with $M_2$ a genus-$h$ Riemann surface, the dominant contribution is disconnected, rather than $K$, when all cycles are sufficiently small.
In this regime, the contribution from $K$ accounts for the subdominant effects of black hole states propagating in the cycles.

Now suppose we follow~\cite{SchWit22} and discussion of Sec.~\ref{sec:largeN} to interpret~\eqref{apar} on the boundary 
as resulting from taking an averaged large $N$ limit, i.e.,  
\be \label{apar1}
F (X', X) = \vev{Z_{M_d} (X) Z_{M_d} (X')} 
\ee
where $Z_{M_d} (X)$ denote the boundary CFT partition function on Euclidean manifold $M_d$ with $X$ insertion and $\vev{\cdots}$ denotes an average over $N$. For definiteness, consider again $d=2$ with $M_2$ a genus-$h$ Riemann surface. 
The contribution to $Z_{M_2}$ from black hole states propagating in its cycles involves a sum over OPE coefficients $f_{ijk}$ among operators $\sO_i$ of dimension $O(N^2)$, where $N$-dependent oscillatory terms can be expected. The bulk saddle $K$ can then be understood as encoding the correlations between $Z_{M_d}(X)$ and $Z_{M_d}(X')$ that arise in an averaged large-$N$ limit.
In this light, the relation implied by~\eqref{overl},\eqref{ove1}, and\eqref{apar1} is especially striking: it links the overlap of ``low-energy'' states in the Lorentzian closed universe to the high-energy behavior of the boundary CFT, governed by black hole states. This is reminiscent of the situation in AS$^2$ cosmology, where the baby universe arises from such oscillatory terms and averages.

The above discussion applies in the regime of large but finite $N$, where the saddle-point approximation can be used to sum over contributions from different saddles.

\section{Conclusions and discussion}\label{sec:Conc} 

\subsection{Immediate future directions for AS$^2$ and MM cosmologies}

In this paper, using the algebraic formulation of AdS/CFT, we have developed a holographic description of certain closed universes~\cite{AntSas23,MalMao04} that arise in AdS. In particular, we identified boundary algebras that encode physical operations within the closed universe. This construction opens new avenues for investigating the structure of quantum gravity in closed cosmologies. Immediate directions for further study include: 

\ben 

\item  Develop a more systematic understanding of the classes of algebras---characterized by their improved UV behavior---that describe closed universes. 

\item The local physics of closed universes can in principle be systematically reconstructed from their holographic descriptions, order by order in the $1/N$ expansion. At leading order, this involves bulk operator reconstruction and correlation functions, while at the next order one can, for example, recover the Casimir energies of closed universes.

\item Understand boundary interpretations of big bang and big crunch singularities. 

\item The formulation here does not rely on introducing observers by hand; rather, it should provide a framework in which time, observers, and relational bulk physics can emerge. For example, in the AS$^2$ cosmology, the mass shell should provide a natural ``observer'' and we can in principle dress observables to it. In MM cosmology, the parent description prior to taking the quotient serves as a useful starting point for understanding the emergence of bulk time.

\een 
 A more detailed examination of these issues is left to future work.

\subsection{Averaged large $N$ limit} \label{sec:avergN}

We proposed that the baby universe in the AS$^2$ cosmology has its boundary origin in $N$-dependent oscillatory terms, and that its boundary description requires an averaged large-$N$ limit. As made evident by the no-go theorem of Gesteau and our discussion, the averaged limit gives rise to richer large-$N$ physics, and thus to richer gravity systems. 

The presence of $N$-dependent oscillatory terms, and their implications for the averaged large-$N$ limit, should be a generic phenomenon. As discussed earlier in~\cite{SchWit22}, averaging over $N$ provides a natural interpretation of wormhole contributions that arise in various contexts. It is therefore important to examine more carefully whether the physical origin of these wormholes can indeed be traced to $N$-dependent oscillatory terms, or whether there exist other concrete mechanisms. In fact, as a support for this interpretation, the second term in~\eqref{Mrenyi} corresponds to a replica-wormhole contribution on the gravity side.

Reinforcing the earlier proposal in~\cite{SchWit22}, we advocated that an averaged large-$N$ limit should be an essential part of the AdS/CFT dictionary.  This notion is conceptually distinct from ensemble averages over different theories: the former encodes universal behavior across different values of $N$ within a single theory; the latter averages over a distribution of genuinely distinct theories.

The need for an averaged large-$N$ limit should have significant implications for our understanding of the AdS/CFT duality. 
 For an observable $f_{N}$, pointwise convergence means that for sufficiently large $N$, $f_{N}$ stays arbitrarily close to its limiting value. By contrast, an averaged limit allows persistent deviations: no matter how large $N$ is, $f_{N}$ may remain far from the semiclassical value suggested by gravity. The boundary quantities should be compared with those on the gravity side only after $N$-average. 
 
 It is therefore of great importance to understand the mathematical nature of the averaged large-$N$ limit. 
 Does there exist a universal averaging procedure across all observables---and, if so, what form it takes.  
To illustrate the challenges for such a procedure,  consider a boundary observable $F_N$ with the schematic asymptotic expansion,  
\begin{equation}
F_N = f_0 + \frac{f_1}{N^2} + \frac{f_2}{N^4} + \cdots 
      + e^{-c N^2} \left( b_0 + \frac{b_1}{N^2} + \cdots \right) 
      + \cos ( d N^2)  + \cdots \, .
\end{equation}
The presence of rapidly oscillating terms such as $\cos (d N^2)$ prevents $F_N$ from having a pointwise large-$N$ limit. 
Ideally, we would like an averaging procedure that removes these oscillatory contributions {\it without} disturbing the non-oscillatory part of the expansion, nor the asymptotic expansions of quantities that already admit a good pointwise large-$N$ limit.
If the procedure does disturb these expansions, it should be in such a way that does not destroy many known successful agreements between boundary and bulk quantities. 

As with any averaging procedure, an $N$-average interacts nontrivially with nonlinear operations. A concrete example arises in the calculation of the R\'enyi entropy in~\eqref{Mrenyi}. There, by discarding the terms involving fluctuating phases, we have effectively performed an $N$-average on the numerator. More appropriately, however, one should first compute $S_2^{(R)}$ in a finite-$N$ theory, perform the $N$-average, and only then take the large-$N$ limit. At present, we are not yet able to carry this out. If one also averages $Z^2$ in the denominator, an additional contribution appears, corresponding to a replica wormhole on the gravity side, as included in~\cite{AntSas23}.

In usual discussions, the large-$N$ limit is implicitly assumed to be pointwise. In tensor-network models of AdS/CFT, this assumption is crucial\footnote{I would like to thank Elliott Gesteau for discussions on this point.}: the code space for the $N=\infty$ theory is embedded into the finite-$N$ Hilbert space, which presupposes a relatively ``unique'' Hilbert space whose physics does not fluctuate wildly at large but finite $N$. Otherwise, the code error cannot be controlled. The no-go theorem of Gesteau suggests that tensor-network models based on a strictly pointwise limit should not capture the baby universe on their own. This, however, does not rule them out; rather, it points to the need for additional elements. Indeed, this may explain why constructions such as those of~\cite{AntSas23,HarUsa25,EngGes25b,AntRat25} need to incorporate random averages and post-selection.

 


\subsection{Toward the mathematical structure of quantum gravity}

The algebraic approach advocated in this paper offers a unified perspective on quantum gravity, independent of asymptotic structures or the sign of the cosmological constant. For closed universes and asymptotically flat spacetimes, rather than searching for a holographic screen to support a boundary theory, one instead seeks to identify the relevant algebras and states. In this framework, asymptotic structures and the sign of the cosmological constant become part of the data specifying a state, rather than essential ingredients in the formulation of the theory. Notions such as the Hilbert space and degrees of freedom emerge as derived, state-dependent concepts. We hope to develop these ideas further in future work.

\subsection{Comparison with other approaches to AS$^2$ cosmology} 

Here we offer some general remarks on alternative approaches that have been proposed to describe the baby universe in AS$^2$ cosmology.  

For this purpose, we first highlight some key questions in the boundary description of the AS$^2$ cosmology: 

\ben 
\item[Q1:] Is the identification of PETS  with the AS$^2$ cosmology correct? 

\item[Q2:] Does the heavy operator matrix elements $\OO_{ab}$ has a well-defined large $N$ limit (other than the non-oscillatory $e^{-\ha N^2 f_0}$ factor)? 

\een

A tensor network model for AS$^2$ cosmology, including a description of the baby universe, was already discussed in~\cite{AntSas23} and further developed in~\cite{AntRat25}. A key element of this construction is the need to perform random averages (or coarse graining) of the heavy operator.
The requirement of coarse graining in order to describe the baby universe implies that the answer to Q1 above is ``no'': AS$^2$ cosmology is not dual to the PETS obtained by inserting a single generic heavy operator, but rather to the state obtained by averaging over a family of heavy operators. This approach also implicitly assumes that the answer to Q2 is ``yes.''
In this coarse-graining framework, there is {\it no} need to modify the AdS/CFT dictionary; what must be refined is the identification between PETS and AS$^2$ cosmology.


In contrast, in the averaged large-$N$ approach advocated in this paper, we assume the answer to Q1 is ``yes'' but the answer to Q2 is ``no,'' motivated by the possible existence of $N$-dependent oscillatory behavior. A negative answer to Q2 then necessitates introducing an averaged large-$N$ limit. This, in turn, requires a modification of the AdS/CFT dictionary and is, in a sense, more radical than the coarse-graining framework.

A third approach~\cite{HarUsa25,EngGes25b}, based on the premise of a one-dimensional Hilbert space for closed universes, introduces a ``classical'' (or ``distinguished'') observer equipped with a large Hilbert space to account for the semiclassical physics of a closed universe.\footnote{Such an interpretation was also implicit in the discussion of~\cite{AntSas23}; see also~\cite{AntRat25}.} In practice, the tensor network model for this approach has the same structure as that of~\cite{AntSas23}.
 Conceptually, however, the ``classical observer'' framework is fundamentally different from both coarse graining and the averaged-$N$ approach.
In this proposal, it is implicitly assumed that the answers to both Q1 and Q2 are ``yes.'' This leaves no room for describing the baby universe within the standard AdS/CFT framework, as implied by Gesteau's no-go theorem. 
To account for the baby universe, the AdS/CFT dictionary is modified by introducing external ``forcing"  such that certain internal observer becomes classical, together with additional rules governing the observer.  This makes it the most radical departure from the standard AdS/CFT framework.


To summarize, clarifying the answers to Q1 and Q2 from the CFT side, together with examining whether closed universes possess one-dimensional Hilbert spaces, will help determine the most appropriate description of the baby universe.


\subsection{One-dimensional Hilbert space for a closed universe?} 
\label{sec:one-dim}


Our holographic description of closed universes within the standard AdS/CFT framework indicates that, for sufficiently small but finite $G_N$, no dramatic departures from semiclassical physics are expected and semiclassical states provide accurate approximations. Several recent works have proposed that the Hilbert space of a closed universe is one-dimensional (see, e.g.,~\cite{McNVaf20,MarMax20,UsaWang24,UsaZha24,HarUsa25,AbdAnt25}).\footnote{A one-dimensional Hilbert space is just $\mathbb{C}$; which is at odds with a macroscopic closed universe populated by stars and galaxies, unless additional structure is introduced.} Here we briefly review the arguments\footnote{For definiteness, we follow the order collected in Sec. 2.1 of~\cite{HarUsa25}.}, highlighting their weaknesses.

\medskip 
 {\it $\bullet$ Entanglement wedge of a reference system} 
\medskip  

The argument~\cite{AlmMah19a,HarUsa25} goes as follows. Suppose a physical system $S$ in a closed universe is entangled with a reference system $Q$. Applying the QES formula to $Q$ would then suggest that the entanglement wedge of $Q$ encompasses the entire closed universe no matter what $Q$ is. This would seem to imply that the closed universe can only have a one-dimensional Hilbert space.

 This conclusion is flawed, as follows directly from our discussion in Sec.~\ref{sec:AEW}, where we presented an explicit example in which the QES formula fails for an $O(N^0)$ entanglement and no entanglement wedge exists. We stress that the situation here is fundamentally different from the Page curve calculation for an evaporating black hole~\cite{Pen19,AlmEng19}, where the entanglement is of order $O(1/G_N)$ and both the QES formula and the usual rules of entanglement wedge reconstruction can be safely applied.

\medskip 
 {\it $\bullet$ Rank of the Gram matrix} 
\medskip  

The argument~\cite{UsaWang24,UsaZha24} (see also~\cite{AbdAnt25} for an explicit calculation in JT gravity) is based on the following {\it topological} toy model~\cite{MarMax20} for closed universes:
\ben
\item The Hilbert space $\sH_{\rm close}$ associated with a closed universe is spanned by states of the form $\ket{i}\in \sH_{\rm close}$, each corresponding to a {\it topological} boundary labeled by an index $i$.

\item The overlap $\vev{i|j}$ is obtained by summing over all possible ways of connecting the topological boundaries associated with $i$ and $j$.
\een
In such a setup, the matrix $M_{ij}=\vev{i|j}$ is necessarily rank-one: the boundary conditions defining $\Tr M^n$ and $(\Tr M)^n$ coincide, and thus the two must be proportional.\footnote{A version of this argument first appeared in~\cite{PenShe19}.}

Toy topological models of this kind are valuable for building intuition, but extending both the setup and its conclusions to realistic systems such as those considered in this paper is far more challenging.
 First, except in certain low-dimensional cases, the rules governing Euclidean path integrals in quantum gravity remain poorly understood beyond the semiclassical regime of saddle-point approximation. It is unclear, for instance, whether Euclidean path integrals can actually be defined properly, and if so, whether all closed-universe states can be generated from operator insertions on some Euclidean boundaries, or what restrictions should be imposed on the path integrals.\footnote{For example, path-integral definitions of closed-universe wave functions may require careful analytic continuation and contour prescriptions~\cite{HarHaw83}. This alone could invalidate the argument for the proportionality of $\Tr M^n$ and $(\Tr M)^n$.} Second, even if one grants that closed-universe states can be defined via a boundary theory on a Euclidean manifold $\sN$ with possible operator insertions, the resulting inner-product matrix need not be well defined outside purely topological settings. For example, even in $d=2$ there exists a functional continuum of possible boundary manifolds on which to place the CFT to define closed-universe wavefunctions, making it unclear how to diagonalize such a functional continuum of states or meaningfully assign a rank.

\medskip 
 {\it $\bullet$ Inner product in a concrete CFT dual} 
\medskip  

This argument~\cite{HarUsa25} considers a setup similar to that of Maldacena-Maoz, discussed in Sec.~\ref{sec:MM}. It asserts that the Euclidean preparation of the state in Sec.~\ref{sec:connE} leads to a factorized inner product if the partition function factorizes. 

As noted in Sec.~\ref{sec:connE}, however, the relation between the inner product in the MM closed universe and the Euclidean path integral is more subtle, so the argument of~\cite{HarUsa25} does not apply.
 
\medskip 
 {\it $\bullet$ Swampland condition} 
\medskip  

It was argued in~\cite{McNVaf20} that the swampland condition forbidding free parameters in quantum gravity rules out $\alpha$-parameters arising from the absorption and emission of baby universes. While a one-dimensional Hilbert space for a closed universe would indeed imply the absence of $\alpha$-parameters, the converse does not necessarily hold: $\alpha$-parameters might also be removed by other dynamical or consistency requirements without reducing the Hilbert space of a closed universe to a single dimension.

\vspace{0.2in}   \centerline{\bf{Acknowledgements}} \vspace{0.2in}
We would like to thank Stefano Antonini, Justin Berman, Netta Engelhardt,  Daniel Harlow, Adam Levine, Juan Maldacena, Martin Sasieta, Brian Swingle, Herman Verlinde, Edward Witten, Ying Zhao, and 
in particular Ping Gao and Elliott Gesteau for discussions. This work is supported by the Office of High Energy Physics of U.S. Department of Energy under grant Contract Number  DE-SC0012567 and DE-SC0020360 (MIT contract \# 578218), and was made possible through the support of grant \#63670 from the John Templeton Foundation.

\appendix

\section{Correlation functions from more general ansatz} \label{app:generalization}

We now calculate correlation functions in PETS using the more general ansatz~\eqref{newsan}, 
\be 
\OO_{ab}  = e^{- \ha N^2 f_0} e^{-{\lam \ov 2} (E_a + E_b)} f^\ha (E_a , E_b ) e^{i N^2 g (E_a , E_b) + i h (E_a, E_b)}  ,
\ee
where for the convenience of later comparing with the gravity results, we have renamed the function $f$ by extracting from it an exponential factor $e^{-{\lam \ov 2} (E_a + E_b)}$ and introduced an order $O(N^0)$ phase $h (E_a, E_b)$. 
We will also relax~\eqref{newsan1} by imposing 
\be \label{hlan}
g_{ab} = g_{cd} , \quad \text{when} \quad E_a - E_b = E_c - E_d ,  
\ee
as we will see below~\eqref{newsan1} could be too stringent. 

Consider first the partition function, which does not depend on the phase, 
\bega 
Z = e^{- N^2 f_0} \sum_{a, b} e^{-\tilde \b_L E_a - \tilde \b_R E_b}  f (E_b , E_a)  , \\ 
\tilde \b_L = \b_L + \lam, \quad \tilde \b_R = \b_R + \lam 
\ . 
\end{gather} 
$\tilde \b_R$ and $\tilde \b_L$ can be identified with the ``renormalized'' inverse temperatures~\cite{AntSas23} on the gravity side due to the presence of the mass shell. 

We can now define $R$ and $L$ density operators by 
\bega 
\ga_R (E) =  {e^{-N^2 f_0}  \ov Z}   e^{-\tilde \b_R   E}  \sum_{a} e^{-\tilde \b_L  E_a }   f(E,  E_a) ,  \\
\ga_L (E) =  {e^{-N^2 f_0}  \ov Z} e^{-\tilde \b_L  E }    \sum_{a}  f(E_a , E)  e^{-\tilde \b_R   E_a}   \ .
\end{gather} 
By definition we have ($H$ is the CFT Hamiltonian) 
\be 
{\rm Tr} \ga_R (H) = \sum_b \ga_R (E_b) = 1, \quad  {\rm Tr} \ga_L  (H) = \sum_b \ga_L (E_b) = 1 \ .
\ee

Now consider correlation functions. We will consider a Hermitian and time-reversal invariant operator $\sO$. We will assume some kind of averaged large $N$ limit such that $N$-dependent phases always cancel. Then, 
\bega 
G_{RR} = \vev{\Psi_\OO|\sO_R (t_1) \sO_R( t_2)|\Psi_\OO} = \vev{\OO^\da(-i \b_R/2) \sO(t_1) \sO(t_2) \OO (i \b_R/2)} \cr
= {1 \ov Z} \sum_{a,b,c,d} e^{-(\b - \b_R)  E_a } \OO^*_{ba}  e^{-(\b_R /2 - i t_1) E_b} 
\sO_{bc} e^{- i (t_1 - t_2) E_c} \sO_{cd} e^{- E_d ( i t_2 +  \b_R/2)} \OO_{da} \\
= {e^{-N^2 f_0}  \ov Z} 
\sum_{a,b,c,d} e^{-\tilde \b_L  E_a }  e^{-(\tilde \b_R /2 - i t_1) E_b} e^{-i N^2 (g_{ba} - g_{da})} e^{ i (h_{da} - h_{ba}) } \cr
\sO_{bc} e^{- i (t_1 - t_2) E_c} \sO_{cd} e^{- E_d ( i t_2 + \tilde \b_R/2)}
f^\ha (E_b,  E_a)  f^\ha (E_d, E_a) 
\end{gather}
Due to the highly fluctuating phase, the dominating contribution to the sum comes from the part where the phase vanishes, i.e., the part with $d =b$. We thus find  
\bega 
G_{RR} ={e^{-N^2 f_0}  \ov Z}  \sum_{a, b,c} e^{-\tilde \b_L  E_a } f(E_b , E_a)  e^{-\tilde \b_R  E_b} 
\sO_{bc} e^{- i (t_1 - t_2) (E_c- E_b)} \sO_{cb} \\
=  \Tr \le(\ga_R (H)  \sO(t_1) \sO(t_2) \ri) \ .
\end{gather}  
We thus conclude that correlation functions of single-trace operators on the $R$-boundary are given by those in the  density operator 
$\ga_R (H)$, and are time-translationally invariant.  This appears to be consistent with the gravity description. 
Similarly, we find 
\be 
G_{LL} = \vev{\Psi_\OO|\sO_L (t_1) \sO_L( t_2)|\Psi_\OO} 
=  \Tr \le(\ga_L (H)  \sO(t_1) \sO(t_2) \ri) \ .
\ee  


Now consider $G_{RL}$, which can be written as an OTOC, 
\bega 
G_{RL} \equiv  \vev{\Psi_\OO|\sO_R (t_1) \sO_L( t_2)|\Psi_\OO} 
= \vev{ \OO^\da (-i \b_R/2) \sO (t_1) \OO (i \b_R/2) \sO(\tilde t_2)}_\b \cr
= {1 \ov Z} \sum_{a,b,c,d} e^{-(\b - \b_R/2 + i \tilde t_2)  E_a } \OO^*_{ba}  e^{-(\b_R /2 - i t_1) E_b} 
\sO_{bc} e^{- i (t_1 - i \b_R/2) E_c} \OO_{cd} e^{E_d ( i \tilde t_2 + \b_R/2)} \sO_{da} \\
= {e^{-N^2 f_0} \ov Z} \sum_{a,b,c,d} 
e^{-(\b/2 + \tilde \b_L/2 + i \tilde t_2)  E_a } e^{-i N^2 (g_{ba} - g_{cd})} e^{i (h_{cd} - i h_{ba})} f^\ha (E_b , E_a)
f^\ha (E_c , E_d) \cr
e^{-(\tilde \b_R /2 - i t_1) E_b} 
\sO_{bc} e^{- i (t_1 - i \tilde \b_R/2) E_c} e^{E_d ( i \tilde t_2 + \b_R/2 - \lam/2)} \sO_{da} 
\label{heq}
\end{gather} 
where $\tilde t_2 = {\b \ov 2} - t_2$. From~\eqref{hlan}, cancellation of the order $O(N^2)$ phases requires 
\be \label{lan1}
E_b - E_a = E_c - E_d \equiv \om  \ .
\ee
We then find that 
\bega 
G_{RL} 
= {e^{-N^2 f_0} \ov Z}  \sum_{a,d, \om} e^{-{\tilde \b \ov 2} (E_a + E_d)} e^{i (E_a - E_d) (t_1+t_2)} e^{-\tilde  \b_R \om}
\sO_{bc} \sO_{da} \cr
\times e^{i h (E_d+\om, E_d)- i h (E_a + \om, E_a)} f^\ha (E_a + \om , E_a)
f^\ha (E_d +\om, E_d) 
 \end{gather} 
 where indices $b,c$ in $\sO_{bc}$ should be understood as expressed through $a,d$ using~\eqref{lan1}. 
 Note that, in contrast to~\eqref{ehn3}, it is no longer factorized between $R$ and $L$. 
 We see that the structure $G_{RL}$ is somewhat complicated, but an interesting feature is that it is only a function of $t_1 + t_2$. 
It would be interesting to check whether at finite $\mu_\OO$ the gravity calculation of $G_{RL}$ does give a function of $t_1+ t_2$ only. 
Suppose we consider~\eqref{newsan1}, i.e., requiring $b=c$ and $a=d$ in~\eqref{heq}, we find that 
\bega 
G_{RL} = {e^{-N^2 f_0} \ov Z} \sum_{a,b} 
e^{- \tilde \b_L E_a - \tilde \b_R E_b}  f (E_b , E_a)
\sO_{bb} \sO_{aa} 
\end{gather} 
which has no time-dependence at all.

\bibliographystyle{jhep}
\bibliography{all}

\end{document}